\documentclass[11pt,a4paper]{article}
\usepackage{amsmath,amsfonts,amssymb,amsbsy}
\usepackage{mathrsfs,latexsym}
\usepackage{graphicx}
\usepackage{xcolor}
\usepackage{booktabs}
\usepackage{multirow}
\usepackage{multicol}
\usepackage{subfig}
\usepackage{placeins}
\usepackage{blindtext}
\usepackage{appendix}
\usepackage{bm}
\usepackage{times}
\usepackage{float}
\usepackage{booktabs}
\usepackage{wrapfig}
\usepackage[utf8]{inputenc} 
\usepackage[T1]{fontenc} 
\usepackage{amsmath,amssymb}
\newcommand{\flowE}{{\mathcal E}}
\newcommand{\ben}{\begin{enumerate}}
\newcommand{\een}{\end{enumerate}}
\newcommand{\bit}{\begin{itemize}}
\newcommand{\eit}{\end{itemize}}
\newcommand{\beq}{\begin{equation}}
\newcommand{\eeq}{\end{equation}}
\newcommand{\bsa}{\begin{subequations}\begin{eqnarray}}
\newcommand{\esa}{\end{eqnarray}\end{subequations}}
\newcommand{\bea}{\begin{eqnarray}}
\newcommand{\eea}{\end{eqnarray}}
\newcommand{\bean}{\begin{eqnarray*}}
\newcommand{\ean}{\end{eqnarray*}}
\newcommand{\nn}{\nonumber \\}
\newcommand{\non}{\nonumber}

\newcommand{\Op}{\mathcal{O}} 
\newcommand{\Ob}{\mathcal{O}} 

\usepackage{alpha}
\usepackage{macros_alpha}

\begin{document}

\title{
\begin{flushright}
\small{
WUB/20-00\\}
\vskip 0.7cm
\end{flushright}
Scale setting for $\Nf=3+1$ QCD}

\collaboration{\includegraphics[width=2.8cm]{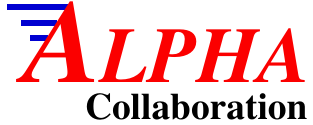}}

\author{Roman H{\"o}llwieser, Francesco Knechtli, Tomasz Korzec}

\address{Dept. of Physics, University of Wuppertal, Gau{\ss}strasse 20, 42119 Germany}

\begin{abstract}
  We present the scale setting for a new set of gauge configurations generated with $\Nf=3+1$ Wilson quarks with a non-perturbatively determined clover coefficient in a massive O($a$) improvement scheme.
  The three light quarks are degenerate, with the sum of their masses being equal to its value in nature
  and the charm quark has its physical mass. 
  We use open boundary conditions in time direction to avoid the problem of topological freezing at small lattice spacings and twisted-mass reweighting for improved stability of the simulations. The decoupling of charm at low energy allows us to set the scale by measuring the value of the low-energy quantity
  $t_0^\star/a^2$, which is the flow scale $t_0$ at our mass point, and comparing it to an $\Nf=2+1$ result in physical units.
We present the details of the algorithmic setup and tuning procedure and give the bare parameters of ensembles with two lattice spacings $a=0.054$fm and $a=0.043$fm. We discuss finite volume effects and lattice artifacts and present physical results for the charmonium spectrum. In particular the hyperfine splitting between the $\eta_c$ and $J/\psi$ mesons agrees very well with its physical value.
\end{abstract}

\maketitle

\tableofcontents

\newpage

\section{Introduction}
\label{sec:intro}

In \cite{Fritzsch:2018kjg} a lattice action was proposed to simulate Quantum Chromodynamics (QCD) with $\Nf=3+1$ quarks on the lattice.
The renormalization and improvement conditions are imposed at finite quark masses. The masses of the up, down and strange quarks are
degenerate and their sum is approximately equal to its value in nature. This not only reduces the number of mass parameters, but also facilitates the simulations not having very light up and down quarks. The mass of the charm quark is close to its physical value.
Quantities like the charmonium spectrum are fairly insensitive to our
  approximation of degenerate light quark masses.
The action in \cite{Fritzsch:2018kjg} was designed to simplify the O($a$) ($a$ is the lattice spacing) improvement pattern for Wilson quarks \cite{Wilson:1975id}
in the presence of a dynamical charm quark. 
The improvement coefficients are determined following the Symanzik improvement program. In a massive scheme for Wilson fermions the
Sheikholeslami--Wohlert coefficient $c_{\rm sw}(g_0^2,aM)$ \cite{Sheikholeslami:1985ij} in the action depends,
besides on the gauge coupling $g_0^2=6/\beta$ also on the quark masses ($aM$ being the bare quark mass matrix). The same is true for the improvement
coefficients of operators. The main result of  \cite{Fritzsch:2018kjg} was a non-perturbative determination of $c_{\rm sw}$ using a finite
volume scheme. In this work we use the action for large volume simulations to test its scaling properties and provide first physical results for the charmonium spectrum. The hyperfine
splitting $m_{J/\psi}-m_{\eta_c}$  between the ground-state vector and pseudoscalar mesons made from two charm quarks is notoriously sensitive to lattice artifacts stemming
from the discretization of the Dirac operator \cite{Choe:2003wx,Liu:2012ze,Cho:2015ffa}, and therefore a good measure of the quality of the action.

One result of our work is the scale setting for $\Nf=3+1$ QCD simulated with the action of
\cite{Fritzsch:2018kjg}.
Scale setting provides a relation between the bare coupling of the simulations and the lattice spacing in fm. The scale
$t_0^\star$ (mass dimension -2) is the flow scale $t_0$~\cite{Luscher:2010iy} at a particular, 
unphysical mass point, namely one where $m_\text{up}=m_\text{down}=m_\text{strange}\equiv m_l$ and
\begin{eqnarray}
   \phi_4 &\equiv& 8t_0\left(m_{\text{K}}^2 + \frac{m_\pi^2}{2} \right) = 12t_0m_{\pi,\text{K}}^2=1.11\, , \label{eq:phi4} \\
   \phi_5 &\equiv& \sqrt{8t_0}\left(m_{\text{D}_{\text{s}}} + 2 m_{\text{D}} \right) = \sqrt{72t_0}m_{\text{D,D}_{\text{s}}} = 11.94 \, . \label{eq:phi5} 
\end{eqnarray}
In our setup the pion and kaon as well as $D$- and $D_s$-mesons are degenerate. In $\Nf=3$ \eq{eq:phi5} plays no role, in our $\Nf=3+1$ simulations discussed here, it is important, that the value of $\phi_5$ corresponds to a charm quark mass, about the same as the one found in nature. 

We measure $t_0^*/a^2$ at a mass point defined by \eq{eq:phi4} and \eq{eq:phi5} and assume that $\sqrt{8t_0^\star} = 0.413(5)(2)$fm to obtain the lattice spacing in fm. This particular value was determined in~\cite{Bruno:2017gxd,Bruno:2016plf} in $\Nf=2+1$ QCD
by determining the ratio of $t_0^{-1/2}$, at the unphysical mass point, with a linear combination of pion and kaon
decay constants at the physical mass point, in the continuum limit.
Relying on a three flavor result is justified by the fact, that when dealing with low-energy quantities like $t_0$, our
$\Nf=3+1$ theory can be well described by an effective theory which to leading order is QCD with just $\Nf=3$ light flavors.
A detailed study of non-perturbative decoupling of the charm quark was performed in a model, QCD with $\Nf=2$ 
mass-degenerate heavy ``charm'' and zero light quarks~\cite{Bruno:2014ufa,Knechtli:2017xgy,Athenodorou:2018wpk,Cali:2019enm}. The effects of the decoupled charm quark at low energies are
incorporated in the matching of the gauge coupling and the (less relevant) quark masses. Power corrections to decoupling are
of the order $(E/m_\text{charm})^2$ and $ (\Lambda/m_\text{charm})^2 $ and stem from higher dimensional terms in the effective Lagrangian.
An important result of the decoupling study was that dimensionless low energy quantities are insensitive to the presence of a dynamical charm quark at our level of accuracy. Indeed, only effects at the permille level were found, which is far below the statistical uncertainty of $\sqrt{8t_0^\star}$.
Should the precision of $\sqrt{8t_0^\star}$ in the $\Nf=3$ theory increase in the future, we might become limited by 
the accuracy to which decoupling holds. A full-fledged simulation program of $\Nf=2+1+1$ QCD, including simulations close to 
the physical mass point, would then be necessary to reduce the scale-setting error further.
We discuss the simulation setup and tuning procedure in section~\ref{sec:setup} and introduce the observables we measure in section~\ref{sec:obs}. In section~\ref{sec:scale} we explain the scale setting of our ensembles with two lattice spacings and discuss systematic errors in section~\ref{sec:sysers}. These ensembles are not only important for a precise determination of the strong coupling constant, but will be very useful for many other physics applications, {\it e.g.} for high precision charm physics. As a first result we therefore present charmonium spectra in section~\ref{sec:spectrum} and extract charmonium masses and hyperfine splitting in good agreement with PDG~\cite{Tanabashi:2018oca} values. We end with concluding remarks and a short outlook in section~\ref{sec:concl}.

\section{Simulations}
\label{sec:setup}

The ensembles were generated using the open-source (GPL v2) package openQCD version 1.6~\cite{openQCD} in plain C with MPI parallelization. This software has been successfully used in various large-scale projects. The measurements of the hadronic correlation functions were carried out with the open-source (GPL v2) program ``mesons''~\cite{mesons}, which is based on various openQCD modules, in particular the openQCD inverters, and hence shares its architecture (C+MPI) and good scalability. 

The simulations are performed on lattices of size $N_t\times N_s^3$. Gauge fields are periodic in spatial directions and absent on temporal links sticking out of the lattice, i.e. we have open boundaries in temporal direction imposed on time slice $0$ and $N_t-1$, hence the physical time extent is $T=(N_t-1)a$ with the lattice spacing $a$.
For a general introduction to lattice QCD simulations with open boundary conditions and twisted-mass reweighting we refer to~\cite{Luscher:2011kk,Luscher:2012av}. 

\subsection{Action}
\label{sec:action}

The full action reads $S=S_{\rm g} + S_{\rm f}$, with a gauge action given by the L\"uscher--Weisz action~\cite{Luscher:1985,Luscher:1985zq} with tree-level coefficients $c_0=5/3$ and $c_1=-1/12$, 
\begin{equation}
   S_{\rm g}[U] = \frac{1}{g_0^2} \left\{ c_0\sum_p w(p){\rm tr}\left[1 - U_p(x) \right] + c_1\sum_r w(r){\rm tr}\left[1-U_r(x)\right]\right\}\, ,
\end{equation}
where the summation is over all oriented plaquettes $p$ and rectangles $r$ (double-plaquettes) on the lattice. 
$U_{p/r}$ is the product of four/six SU(3) gauge fields $U_\mu(x)$ around the elementary plaquette $p$ or rectangle $r$.
The free parameter of the gauge action is the bare coupling $g_0^2 \equiv 6/\beta$ and the weights $w(p)=w(r)=1$ except for spatial plaquettes and rectangles at $T=0$ and $T=(N_t-1)a$ where $w(p)=w(r)=1/2$, i.e. the coefficient of the gluonic boundary improvement term is set to
its tree level value  $c_G = 1$. We do the same with the coefficient of the fermionic boundary improvement term $c_F = 1$.
The fermion action  then reads
\begin{equation}
S_\mathrm{f}[U,\overline \psi,\psi]=
a^4 \sum_{f=1}^4 \sum_x\, \overline \psi_f(x)\, [D_\mathrm{W}+m_{f}]\, \psi_f(x) + S_{\rm SW}\,,
\label{eq:ferm}
\end{equation}
with the  Wilson Dirac operator~\cite{Wilson:1974sk} 
\begin{equation} D_\text{W} = 
\frac{1}{2}\sum_{\mu=0}^3 \{\gamma_\mu(\nabla^*_\mu+\nabla_\mu) -a\nabla^*_\mu\nabla_\mu\} \,,
\end{equation}
where $\nabla_\mu$ and $\nabla_\mu^*$ are the covariant
forward and backward derivatives, respectively, and the Sheikholeslami--Wohlert action $S_{\rm SW}$
in \eq{eq:ferm} is needed for
$\rm{O}(a)$ improvement~\cite{Sheikholeslami:1985ij}.

Simulating a physical charm quark results in lattice artifacts due to large cutoff effects of order proportional to a large value of the lattice charm quark mass $(am_c\approx0.5)$.
In a mass independent scheme the mass dependent O($a$) lattice artifacts are cancelled by
improvement terms where a mass independent coefficient multiplies $aM$ \cite{Luscher:1996sc,Bhattacharya:2005rb}.
Some of these coefficients are only known to one loop in perturbation theory which
may be acceptable for the light quarks but not if the quark mass is large.
Therefore a massive renormalization scheme with close to realistic charm mass $m_c$ and Symanzik improvement for Wilson quarks was proposed~\cite{Fritzsch:2018kjg}. 
This results in mass-dependent renormalization parameters for coupling and quark masses 
as well as a mass-dependent coefficient in the clover action 
\bean
S_{\rm SW} &=& a^5 c_{\mathrm{sw}}(g_0^2,aM) \sum_x \bar{\psi}(x) \frac{i}{4} \sigma_{\mu\nu} \hat{F}_{\mu\nu}(x) \psi(x)\,,
\ean
($\hat F_{\mu\nu}$ is the standard discretization of the field strength tensor~\cite{Luscher:1996sc}), which we determine using the non-perturbative fit formula obtained in~\cite{Fritzsch:2018kjg} 
\begin{equation}
   c_{\rm sw}(g_0^2,aM) = \frac{1+A g_0^2+B g_0^4}{1+(A-0.196)g_0^2}\,,\quad A = -0.257 \,,\quad B = -0.050\,. \label{eq:csw}
\end{equation}

While in a full massive scheme the renormalization and improvement factors would depend on all quark masses, in
practice a hybrid approach is more feasible, where the dependence on the heavy quarks (just the charm quark in our case) is
absorbed into the factors, but light quarks are treated like in a mass-independent scheme. This allows to use the
same value of $c_{\mathrm{sw}}$ along chiral trajectories with constant charm quark mass, while at the same 
time avoiding large $O(am_c)$ artifacts. In fact, \eq{eq:csw} can be used whenever up, down and strange quarks are light,
and the charm quark has close to its physical mass.
In this work we describe simulations at the $SU(3)$ flavor symmetric point, where $M = \mathrm{diag}(m_l, m_l, m_l, m_c )$\footnote{Often, we quote the hopping parameters $\kappa_f=1/(2am_f+8)$ instead of the bare quark masses.}.

\subsection{Algorithms}
\label{sec:algo}

The generation of these configurations proceeds according to a variant of the Hybrid Monte-Carlo (HMC) algorithm~\cite{Duane:1987de}. The classical equations of motion are solved numerically for trajectories of length $\tau = 2$ in all simulations, leading to Metropolis proposals which are accepted with an acceptance rate $\langle P_{acc}\rangle$. In order to reduce the computational cost and obtain a high acceptance rate, we split the action and corresponding forces as explained in detail in this section.  

Since the Wilson Dirac operator is not protected against eigenvalues below the quark mass, 
we use the second version of twisted mass reweighting, suggested in \Ref{Luscher:2008tw}, 
for the asymmetric even--odd preconditioned~\cite{DeGrand:1988vx} Dirac operator 
\beq
\hat D=D_{ee}-D_{eo}(D_{oo})^{-1}D_{oe}\,\qquad D=D_W+m_0=\bigg(\begin{array}{cc}D_{ee}&D_{eo}\\D_{oe}&D_{oo}\end{array}\bigg).
\eeq
The u/d quark doublet\footnote{The openQCD code~\cite{openQCD} is specialized in simulating a light quark doublet with twisted mass reweighting and further quarks with higher masses using the RHMC algorithm~\cite{Kennedy:1998cu,Clark:2006fx}, that's why we treat  the u/d doublet and strange quark separately, even though they are degenerate.} 
is then simulated with a weight proportional to
\begin{equation}
   \text{det}[D^\dagger D ]\rightarrow\text{det}[(D_{oo})^2 ]\text{det}\dfrac{\hat D^\dagger \hat D + \mu_0^2}{\hat D^\dagger \hat D + 2\mu_0^2}\text{det}[\hat D^\dagger \hat D + \mu_0^2]\label{eq:doublet}
\end{equation}
where we introduced an infrared cutoff by the twisted mass $\mu_0$. To further reduce the fluctuations in the forces, we factorize the last term in \eq{eq:doublet} according to~\cite{Hasenbusch:2001ne,Hasenbusch:2002ai,Schaefer:2012tq}
\bean
\det[\hat D^\dagger \hat D+\mu_0^2] = \det[\hat D^\dagger \hat D + \mu_N^2] \times \frac{\det[\hat D^\dagger \hat D + \mu_0^2]}{\det[\hat D^\dagger \hat D + \mu_1^2]}
   \times \ldots \times \frac{\det[\hat D^\dagger \hat D+\mu_{N-1}^2]}{\det[\hat D^\dagger \hat D + \mu_{N}^2]},
\ean
with $a\mu_i$ given by $\{0.0005, 0.005, 0.05, 0.5\}$ for all our ensembles.
The individual factors can be represented by pseudo-fermions such that the combination of twisted-mass reweighting and factorization leads to a pseudo-fermion action for the light quark doublet with $N+2=6$ terms
\begin{equation}
\begin{split}
S_\mathrm{ud, eff}[U,\phi_0,\dots,\phi_{N+1}]=&
\big (\phi_0, \frac{\hat D^\dagger \hat D+ 2\mu_0^2}{\hat D^\dagger \hat D + \mu_0^2}  \phi_0 \big)
+
\sum_{i=1}^N 
\big(\phi_i, \frac{\hat D^\dagger \hat D + \mu_{i}^2 }{\hat D^\dagger \hat D + \mu_{i-1}^2 }  \phi_i \big )
\\
&
+ \big \{ \big (\phi_{N+1}, \frac{1}{\hat D^\dagger \hat D + \mu_N^2} \phi_{N+1} \big )
-2 \log \det  D_\mathrm{oo} \} \,.
\label{eq:effmf}
\end{split}
\end{equation}
where $\phi_i$ are six pseudo-fermion fields with support on the even sites of the lattice.
The u/d quark doublet comes with a reweighting factor 
\beq
W_{ud}=\text{det}[\hat D^\dagger \hat D(\hat D^\dagger \hat D + 2\mu_0^2)(\hat D^\dagger \hat D + \mu_0^2)^{-2}]\,,\label{eq:rwdblt}
\eeq
which is estimated stochastically. 

The strange and charm quarks are simulated with the RHMC algorithm~\cite{Kennedy:1998cu,Clark:2006fx}, decomposing
\beq
\text{det}\hat D_q = W_q\text{det}R_q^{-1}\,,\qquad q\in[s,c]
\eeq
into reweighting factors (to be estimated stochastically again)
\beq
W_q=\text{det}[\hat D_q R_q]\label{eq:rwrat}
\eeq
and Zolotarev optimal rational approximations of $(\hat D_q^\dagger\hat D_q)^{-1/2}$~\cite{Achiezer:1992} 
\beq
R_q=A_q\prod_{k=1}^{N_p^q}\dfrac{\hat D_q^\dagger \hat D_q+\nu_{q,k}^2}{\hat D_q^\dagger \hat D_q + \omega_{q,k}^2}\label{eq:zrat}
\eeq
in the spectral ranges $[r_a,r_b]^q$ of $\hat D_q^\dagger\hat D_q$ with a number of poles $N_p^q$, optimized during the tuning process, which uniquely determine the parameters $A_q, \omega_q$ and $\nu_q$. 
The Zolotarev rational function \eq{eq:zrat}, in the case of the strange quark, is further broken into several factors, introducing 
separate pseudo-fermion fields for the five smallest $\omega_{s,k},\nu_{s,k}$, whereas the determinant of the remaining factors (first seven poles) is expressed as a single pseudo-fermion integral. In practice, this product is rewritten via partial fraction decomposition into a sum 
\bea
\prod_{k=1}^{7} 
\frac{\hat D_s^\dagger \hat D_s + \nu_{s,k}^2}{\hat D_s^\dagger \hat D_s + \omega_{s,k}^2}=1+\sum_{k=1}^{7} 
\frac{\rho_{s,k}}{\hat D_s^\dagger \hat D_s + \omega_{s,k}^2}
\eea
with
\bea
\rho_{s,k}=(\nu_{s,k}^2-\omega_{s,k}^2)\prod_{m=1,m\neq k}^7\frac{\nu_{s,m}^2-\omega_{s,k}^2}{\omega_{s,m}^2-\omega_{s,k}^2}\,.
\eea
The effective action for the strange quark with $N_p^s=12$ poles reads
\bea
S_\mathrm{s, eff}[U,\phi_7,\dots,\phi_{12}]
&=&\big(\phi_{7}, (1+\sum_{k=1}^{7} 
\frac{\rho_{s,k}}{\hat D_s^\dagger \hat D_s + \omega_{s,k}^2}) \phi_{7}\big)\\
&&-\log \det D_\mathrm{oo}^s+
\sum_{l=8}^{12}
\big(\phi_{l}, \frac{\hat D_s^\dagger \hat D_s + \nu_{s,l}^2}{\hat D_s^\dagger \hat D_s + \omega_{s,l}^2}
\phi_{l}\big).\quad
\label{eq:effrhms}
\eea
The charm quark contribution to the action is not further factorized
\bea
S_\mathrm{c, eff}[U,\phi_{13}]=&
\big(\phi_{13}, (1+\sum\limits_{k=1}^{N_p^c}
\frac{\rho_{c,k}}{\hat D_c^\dagger \hat D_c + \omega_{c,k}^2}) \phi_{13}\big)
-\log \det D_\mathrm{oo}^c\,,
\label{eq:effrhmc}
\eea  
such that we have 13 pseudo-fermion fields and 14 actions in total, with forces summarized in \tab{tab:param}. 
The molecular dynamics equations are integrated using multi-level higher order (2nd and 4th order OMF) integrators~\cite{Omelyan:2003}. Linear systems involving the Dirac operators are dealt with using low-mode-deflation~\cite{Luscher:2007se} and SAP-preconditioned~\cite{Luscher:2003qa} Krylov space solvers based on local coherence~\cite{Luscher:2007es,Frommer:2013fsa}, the tuning of corresponding parameters is discussed in the next section.

\subsection{Parameter tuning}

For the algorithmic parameters, we started with the setup of CLS's H400 simulation, cf.~\cite{Bruno:2014jqa}, to which we added the charm quark. The new contribution to the action was not further factorized and the corresponding forces were integrated on the intermediate level of our three level integrator. The gauge fields are integrated on the innermost level with the fourth order integrator suggested by Omelyan, Mryglod, and Folk (OMF)~\cite{Omelyan:2003}, most of the fermion forces are on the intermediate level (4th order OMF) and only the most expensive doublet factor (the one with $\phi_0$ in \eq{eq:effmf}) and four largest poles of the strange rational function are on the outermost (coarsest) level using a second order OMF. 
For most fermion forces we use the locally deflated solver~\cite{Luscher:2007se,Luscher:2007es,Frommer:2013fsa} with a deflation subspace block size $4^4$ and 24 deflation modes per block. We set the relative residua to $10^{-12}$ in the acceptance-rejection step and to $10^{-11}$ in the force computation. The parameters of the rational functions, i.e. number of poles and spectral ranges, are adjusted during thermalization by calculating the reweighting factors $W_{s,c}$ and the true spectral range of $|\gamma_5\hat D|$ for a subset of the generated gauge field configurations. The simulation and algorithmic parameters are summarized in \tab{tab:param}.

\section{Observables}\label{sec:obs}

\subsection{Wilson-flow observables}
The Wilson flow can be used to define quantities with a finite continuum limit in lattice QCD, starting out with a smoothing flow equation \cite{Luscher:2010iy,Narayanan:2006rf}
\begin{equation}
   \partial_t V_\mu(x,t) = -g_0^2 \left\{\partial_{x,\mu} S_{\rm W}[V] \right\} V_\mu(x,t)
   ,\qquad V_\mu(x,0) = U_\mu(x)\, ,
\end{equation}
with smeared gauge field $V_\mu(x,t)$ at flow time $t$ and the corresponding Wilson gauge action $S_{\rm W}=\frac{1}{g_0^2}\sum_p {\rm tr}\left[1 - V_p(x) \right]$.

Using a symmetrized clover definition of the field strength tensor $\hat G_{\mu\nu}(x,t)$ constructed from the smooth fields $V_\mu(x,t)$, the time slice action density $\flowE(x_0,t)$ and the global topological charge $Q(t)$ can be defined as
\bea
\flowE(x_0,t)&=-\frac{a^3}{2L^3}\sum_{\vec x}  \mathrm{tr} \{ \hat G_{\mu\nu}(x,t) \,  \hat G_{\mu\nu}(x,t) \} \,,\label{eq:E}\\
Q(x_0,t)&=-\frac{a^4}{32\pi^2}\sum_{\vec x} \epsilon_{\mu\nu\alpha\beta}\, \mathrm{tr} \{ \hat G_{\mu\nu}(x,t) \,  \hat G_{\alpha\beta}(x,t) \} \,.
\label{eq:Q}
\eea
Since significant boundary effects in $\flowE(x_0,t)$ and $Q(x_0,t)$ were observed in \cite{Bruno:2014jqa},  
we take a plateau average in a range of $x_0$ values away from the boundaries (given in \tab{tab:meas}) to define the vacuum expectation value of the energy $\langle \flowE(t) \rangle$ ant topological charge $Q(t)$. 
Correlators constructed from gauge fields at $t>0$ are automatically renormalized \cite{Luscher:2011bx}, which allows to define low-energy scales $t_0$\cite{Luscher:2010iy}, $t_c$ and $w_0^2$\cite{Borsanyi:2012zs} as the flow times $t$ at which 
\begin{equation}
   t^2 \langle \flowE(t) \rangle = 0.3,\;0.2\quad\mbox{and}\quad t^2 \langle\partial \flowE(t)/\partial t \rangle = 0.3\label{eq:t0},
\end{equation}
respectively, the partial derivative in the case of $w_0$ is evaluated numerically.

\subsection{Meson correlators and masses}

We want to extract meson masses from the zero momentum correlation functions
\begin{equation}
      f_{\Op\Op}(x_0,y_0) = a^6 \sum_{\mathbf{x},\mathbf{y}} \langle \Op(x) \Op^\dagger(y) \rangle 
\end{equation}
The operators $\Ob$ in \tab{tab:meson_states} are of the form $\bar q_1\Gamma q_2$, where $\Gamma$ is $4\times4$ spin matrix. 
Integrating out the fermions leaves us with a single, connected diagram of the form
\bea
 -\sum_{\boldsymbol x,\boldsymbol y} 
   \left\langle {\rm tr}\left[ 
   \Gamma \mathcal{S}_2(x,y) \bar\Gamma \mathcal{S}_1(y,x)
   \right] \right\rangle^{\rm gauge}\;.
\eea
with $\bar \Gamma \equiv \gamma_0 \Gamma^\dagger\gamma_0$. The trace acts in color and spin space and the propagators $\mathcal{S}_i$ are given by the inverse Dirac operator
\begin{equation}
   \sum_y [D_W(x,y) + m_i] \mathcal{S}_i(y,z) = \delta_{x,z}\, .
\end{equation}

\begin{table}[ht!]
\begin{center}
\renewcommand{\arraystretch}{1.4}
\renewcommand{\tabcolsep}{8pt}
\begin{tabular}{lllll}
\hline
State        & $J^{PC}$ & $\Ob$ & Particle & Plateau A1, A2, B              \\
\hline
Scalar       & $0^{++}$ & $S=\bar{c}c'$ & $\chi_{c_0}$ & \{25-43\}, \{25-43\}, \{26-43\} \\
Pseudoscalar & $0^{-+}$ & $P=\bar{u}\gamma_5d$ & $m_\pi$ & \{30-70\}, \{30-90\}, \{40-100\}  \\
& & $P=\bar{u}\gamma_5s$ & $m_{\text{K}}$ & \{30-70\}, \{30-90\}, \{40-100\} \\
& & $P=\bar{u}\gamma_5c$ & $m_{\text{D}}$ & \{30-70\}, \{30-70\}, \{40-80\} \\
& & $P=\bar{s}\gamma_5c$ & $m_{\text{D}_{\text{s}}}$ & \{30-70\}, \{30-70\}, \{40-80\} \\
& & $P=\bar{c}\gamma_5c'$ & $\eta_c$ & \{30-70\}, \{30-70\}, \{40-80\} \\
Vector       & $1^{--}$ &$V_{i}=\bar{c}\gamma_{i}c'$ & $J/\psi$ & \{30-70\}, \{30-80\}, \{40-100\} \\
Axial vector & $1^{++}$ & $A_{i}=\bar{c}\gamma_{i}\gamma_{5}c'$ & $\chi_{c_1}$ &\{22-35\}, \{22-35\}, \{25-35\} \\
Tensor\footnotemark  & $1^{+-}$ & $T_{ij}=\bar{c}\gamma_i\gamma_jc'$ & $h_c$ & \{18-25\}, \{18-25\}, \{22-30\} \\
\hline
\end{tabular}
\end{center}
\caption{Meson state interpolators and particle names which are the closest relatives in nature.}\label{tab:meson_states}
\end{table}
\footnotetext{The notation refers to the $\gamma$-structure of the operator.}

An efficient way to carry out the calculation is to use stochastic sources, we use 32 noise vectors per time-slice, which amounts to 32 inversions per $y_0$ value and Dirac structure. 
In Table~\ref{tab:meson_states} we list the meson interpolators for the various states investigated and their particle names which are the closest relatives in nature.
An improved signal and exact symmetries are achieved by defining the averages
\begin{eqnarray}
\bar f_{PP}(x_0-a) &\equiv& \frac{1}{2} \left( f_{PP}(x_0,a) + f_{PP}(T-x_0,T-a)  \right)\, , \label{eq:avcorr1} \\
   \bar f_{AA}(x_0-a) &\equiv& \frac{1}{2} \left( f_{AA}(x_0,a) + f_{AA}(T-x_0,T-a)  \right)\, , \\
  \bar f_{VV}(x_0-a) &\equiv& \frac{1}{6} \sum_{k=1}^3  \left( f_{V_kV_k}(x_0,a) + f_{V_kV_k}(T-x_0,T-a)  \right) \, , \\
  \bar f_{SS}(x_0-a) &\equiv& \frac{1}{2} \left( f_{SS}(x_0,a) + f_{SS}(T-x_0,T-a)  \right)\, , \\
  \bar f_{TT}(x_0-a) &\equiv& \frac{1}{6} \sum_{j>i} \left( f_{T_{ij}T_{ij}}(x_0,a) + f_{T_{ij}T_{ij}}(T-x_0,T-a)  \right).\label{eq:avcorr5}
\end{eqnarray}

The meson masses are extracted from the exponential decay of these correlators at $x_0 \gg a $ via weighted plateau averages 
\begin{equation}
	m = \sum\limits_{x_0 = t_{\rm low}}^{t_{\rm high}} w(x_0+a/2) m^{\rm eff}(x_0+a/2)\bigg/\sum\limits_{x_0 = t_{\rm low}}^{t_{\rm high}} w(x_0+a/2)\, ,
\label{eq:plateau}
\end{equation}
of the effective masses
\begin{equation}
   a m^{\rm eff}(x_0+a/2) \equiv \ln\left(\frac{f(x_0)}{f(x_0+a)} \right)\, .\label{eq:effmass}
\end{equation}
and weights $w$ given by their inverse squared errors. Table~\ref{tab:meson_states} also lists the plateau ranges for the different 
ensembles, chosen such that excited state contributions are negligible.

\subsection{Reweighting, mass derivatives and tuning corrections}\label{sec:rmd}
With reweighting, QCD expectation values are obtained from the simulated ones via
\begin{equation}
   \langle \Ob\rangle_{QCD} = \frac{\langle \Ob\, W\rangle}{\langle W\rangle} \, ,\label{eq:rw}
\end{equation}
with corresponding weights $W=W_{ud}W_sW_c$ given by the product of reweighting factors \eq{eq:rwdblt} and \eq{eq:rwrat} estimated stochastically. We are interested in functions of primary observables
\begin{equation}
   f(\langle \Ob_1\rangle_{QCD},\ldots , \langle \Ob_N\rangle_{QCD},m)
\end{equation}
and their quark mass derivatives $df/dm$. Therefore we first reweight our primary observables according to \eq{eq:rw} and calculate their reweighted mass derivatives via~\cite{Bruno:2016plf}
\begin{eqnarray}
   \frac{d\langle\Ob_i\rangle_{QCD}}{dm} &=& \left\langle \frac{\partial \Ob_i}{\partial m} \right\rangle_{QCD}
                   - \left \langle \Ob_i \frac{\partial S}{\partial m}\right\rangle_{QCD} 
                   + \langle \Ob_i\rangle_{QCD}\left\langle\frac{\partial S}{\partial m}\right\rangle_{QCD}\nonumber  \\
                 &=& \frac{\left\langle \frac{\partial \Ob_i}{\partial m} W \right\rangle}{\langle W\rangle}
                   -\frac{\left \langle \Ob_i \frac{\partial S}{\partial m}W\right\rangle}{\langle W\rangle}
                   +\frac{\langle \Ob_i W\rangle \left\langle\frac{\partial S}{\partial m}W\right\rangle}{\langle W\rangle^2}\,, \label{eq:Wick}
\end{eqnarray}
where $\frac{\partial S}{\partial m_q}=\sum_x \bar{q}(x) q(x)$ gives rise to
$\sum_x {\rm tr}\left[\mathcal{S}(x,x)\right]$ after integration over the quark fields in the third term of \eq{eq:Wick} and also in the second term, if no other
quark fields are present in the observable $\Ob_i$.
This trace is evaluated stochastically using 32 U(1) random sources per configuration. When $\Ob_i$
contains fermionic fields, the second term in \eq{eq:Wick} produces additional Wick contractions
that depend on $\Ob_i$. For the case of the meson correlators these are worked out and measured.
For derived observables $f$, we then use the chain rule
\begin{equation}
   \frac{d\,f(\langle \Ob_1\rangle_{QCD},\ldots,  \langle \Ob_N\rangle_{QCD}, m)}{dm} =
   \sum_{i=1}^N \frac{\partial f}{\partial \langle \Ob_i\rangle_{QCD}}\, \frac{d\langle \Ob_i\rangle_{QCD}}{dm} + \frac{\partial f}{\partial m} \, .
\end{equation}
The partial derivatives can be calculated analytically or estimated numerically
\begin{equation}
   \frac{\partial f}{\partial \langle \Ob_i\rangle_{QCD}} \approx \left[f\left(\ldots, \frac{\langle \Ob_iW\rangle}{\langle W\rangle}+h,\ldots\right) 
                                               - f\left(\ldots, \frac{\langle \Ob_iW\rangle}{\langle W\rangle}-h,\ldots\right)\right]/(2h)
\end{equation}
where a suitable $h$ is obtained from the fluctuations of $\Ob_i$, and we do not have explicit derivatives $\frac{\partial f}{\partial m}$ for the derived observables considered here.

In order to correct for mis-tunings of $\phi_4$ \eq{eq:phi4} and $\phi_5$ \eq{eq:phi5}, we solve the following system of linear equations to get the mass shifts $\Delta m_l$ and $\Delta m_c$ for light and charm quarks
\begin{eqnarray}
   1.11&=&\phi_4+\left(\dfrac{d\phi_4}{dm_u}+\dfrac{d\phi_4}{dm_d}+\dfrac{d\phi_4}{dm_s}\right)\Delta m_l+\dfrac{d\phi_4}{dm_c}\Delta m_c \nn
   11.94&=&\phi_5+\left(\dfrac{d\phi_5}{dm_u}+\dfrac{d\phi_5}{dm_d}+\dfrac{d\phi_5}{dm_s}\right)\Delta m_l+\dfrac{d\phi_5}{dm_c}\Delta m_c \non
\end{eqnarray}
and use these to correct arbitrary observables $f$ accordingly
\begin{eqnarray}\label{eq:shift}
	f_{\rm corrected}&=&f+\left(\dfrac{df}{dm_u}+\dfrac{df}{dm_d}+\dfrac{df}{dm_s}\right)\Delta m_l+\dfrac{df}{dm_c}\Delta m_c\,.
\end{eqnarray}

\section{Scale setting}
\label{sec:scale}

\begin{table}[t]
\centering
\begin{tabular*}{\textwidth}{c @{\extracolsep{\fill}}cccccccc}
\toprule  
ens. & $\frac{T}{a} \times \frac{L^3}{a^3}$ &$\beta$ & $\kappa_l$ & $\kappa_c$ & $a$[fm] & $Lm_\pi^\star$ & MDUs & $\tau_{\rm exp}$\\
\midrule
A0 & $96 \times 16^3$ & 3.24 & 0.13440733 & 0.12784 & 0.054 & 1.77 & 2800 & -  \\
\midrule
A1 & $96 \times 32^3$ & 3.24 & 0.13440733 & 0.12784 & 0.0536(11) & 3.651(13) & 7816 & 80 \\
\midrule
A2 & $128 \times 48^3$ & 3.24 & 0.13440733 & 0.12784 & 0.0538(11) & 5.333(14) & 7736 & 100 \\
	&				&	& 0.134396(14)	& 0.12798(19) & 0.0536(11) & 5.354(13) &	\\
\midrule
B & $144 \times 48^3$ & 3.43 & 0.13599 & 0.13088 & 0.0429(8) & 4.301(19) & 8000 & 160 \\
    &				&		& 0.13599(1) & 0.13090(33) & 0.0428(7) & 4.282(14) \\
\bottomrule
\end{tabular*}
 \caption{Simulation parameters, lattice sizes and statistics of the three ensembles, the second row values for ensembles A2 and B are shifted to the correct tuning points $\phi_4$ and $\phi_5$, using the strategy described in \sect{sec:rmd}, \eq{eq:shift}.}
 \label{tab:sim}
\end{table}

For starting parameters we choose a bare coupling $\beta = 3.24$, light quark masses given by $\kappa_{u,d,s} = 0.134484$ and a charm quark mass by $\kappa_c = 0.12$, obtained from a rough interpolation of the simulation parameters in~\cite{Fritzsch:2018kjg}.
We thermalize on spatially smaller lattices and subsequently double the spatial dimensions, using eight copies of the smaller lattices as the new initial configurations. Unfortunately, the smaller lattices are not helpful for parameter tuning, since we are dealing with large finite volume effects, as discussed in \sect{sec:finV}. We measured the flow observables and meson masses on a more-or-less thermalized subset of configurations of the final lattice size to evaluate $\phi_4$ and $\phi_5$. With the starting parameters the tuning point was missed by quite a bit, therefore, we simulate several other mass points and calculate the derivatives of $\phi_4$ and $\phi_5$ with respect to bare quark masses, in order to gradually reach the correct tuning point. This is non-trivial, since the mass dependence of $t_0$ and the meson masses in $\phi_4$ and $\phi_5$ go in opposite directions, see~\tab{tab:sigmA2} and~\ref{tab:sigmaB} for results on the ensembles A2 and B. With the final parameters 
$\kappa_{u,d,s} = 0.13440733$ and 
$\kappa_c = 0.12784$,
at $\beta=3.24$ we produced two high statistics ensembles A1 and A2 with two different lattice sizes, and a smaller ensemble A0 on an even smaller volume to study finite size effects. Assuming decoupling of the charm quark, {\textit i.e.}, $t_0^\star|_{N_f=3+1}=t_0^\star|_{\Nf=3}+O(1/m_\mathrm{charm}^2)$, our value of $t_0/a^2\approx7.4$ corresponds to a lattice spacing $a \approx 0.054$ fm, where we used the $\Nf=3$ result $\sqrt{8t_0^\star} = 0.413(5)(2)$fm from~\cite{Bruno:2017gxd,Bruno:2016plf}. This makes our smaller volume of ensemble A1 ($L/a=32$) $L\approx 1.73$ fm with $m_\pi L = 3.5$, which is a bit small, but finite size effects seem to be under control, as the comparison with $L/a=48$ shows, see further \sect{sec:finV}. In a next step we tuned an ensemble B at a finer lattice spacing on a $144\times48^3$ lattice with a bare coupling $\beta=3.43$ using the same procedure as described above, yielding final parameters 
$\kappa_{u,d,s} = 0.13599$ and 
$\kappa_c = 0.13088$,
and a lattice spacing $a\approx0.043$fm. This gives a physical lattice extent $L\approx2.06$fm and $m_\pi L = 4.3$.

\begin{table}[h]
\centering
\begin{tabular*}{\textwidth}{c @{\extracolsep{\fill}}ccccccccc}
\toprule  
ens. & plat. & $t_0/a^2$ & $\tau_{int,t_0}$ & $t_c/a^2$ & $\tau_{int,t_c}$ & $w_0^2/a^2$ & $\tau_{int,w_0}$ & $Q^2$ & $\tau_{int,Q^2}$\\
\midrule
A0 & [22:74] & 8.83(23) & 10(2) & 4.12(9) & 9(4) & - & - & 0.83(11) & 6(1) \\ 
\midrule
A1 & [22:74] & 7.42(4) & 16(7) & 3.88(1) & 11(4) & 10.25(12) & 24(12) & 1.13(4) & 5(1) \\ 
\midrule
A2 & [22:106] & 7.37(2) & 27(15) & 3.86(1) & 17(11) & 10.15(6) & 25(13) & 6.55(18) & 3(1) \\
      &          & 7.43(5) & 	       & 3.89(2) &  	    & 10.27(11) &     	  & 6.77(28) & \\
\midrule
B & [30:114] & 11.60(6) & 35(21) & 6.00(2) & 32(19) & 16.57(17) & 29(17) & 1.63(7) & 8(3) \\
   & 		& 11.62(9) & 		& 6.01(3) & 	   & 16.63(25) & 	    & 1.61(9) \\
\bottomrule
\end{tabular*}
 \caption{Flow measurement results and topological charge with integrated autocorrelation times. The second column specifies the plateau range of $x_0$ values where $\flowE(x_0,t)$ and $Q(x_0,t_0)$ are averaged. The second row values for ensembles A2 and B are shifted to the correct tuning points $\phi_4$ and $\phi_5$, using the strategy described in \sect{sec:rmd}, \eq{eq:shift}.}
 \label{tab:meas}
\end{table}

\begin{table}[h]
\centering
\begin{tabular*}{\textwidth}{c @{\extracolsep{\fill}}cccccccc}
\toprule  
ens. & $N_{\rm ms}$ & $am_{\pi,K}$& $\tau_{int}$ & $am_{D,D_s}$& $\tau_{int}$ & $\phi_4$ & $\phi_5$ \\
\midrule
A0 & 700 & 0.310(6) & 2.74(92) & 0.614(17) & 2.24(63) & 10.22(90) & 15.48(43) \\ 
\midrule
A1 & 1954 & 0.1141(12) & 3.86(82) & 0.5232(12) & 1.75(39) & 1.161(22) & 12.098(36) \\ 
\midrule
A2 & 1934 & 0.1111(4) & 4.17(90) & 0.5234(4) & 1.87(34) & 1.092(6) & 12.058(17) \\
	&	& 0.1115(3)  &		& 0.5160(16)	&		& 1.11	&11.94	\\
\midrule
B & 2000 & 0.0896(5) & 6.59(81) & 0.4135(7) & 1.87(17) & 1.116(12) & 11.950(30) \\
	&	& 0.0892(4) &		& 0.4128(17)&			& 1.11	&11.94	\\
\bottomrule
\end{tabular*}
 \caption{Light meson masses and tuning results, the second row values for ensembles A2 and B are shifted to the correct tuning points $\phi_4$ and $\phi_5$, using the strategy described in \sect{sec:rmd}.}
 \label{tab:mass}
\end{table}

Simulation parameters are summarized in Table~\ref{tab:sim}, flow measurement and topological charge results are presented in Table~\ref{tab:meas} and corresponding masses and tuning results are shown in Table~\ref{tab:mass}. The second row values for ensembles A2 and B are shifted to the correct tuning points $\phi_4$ and $\phi_5$, using the strategy described in \sect{sec:rmd}. The corresponding quark mass shifts $a\Delta_{m_l}$ and $a\Delta_{m_c}$ are given in Table~\ref{tab:shifts} together with the final lattice spacings.

\begin{table}[h]
\centering
\begin{tabular}{rccc}
\toprule  
ens.  & $a\Delta_{m_l}$ & $a\Delta_{m_c}$ & $a$[fm]\\
\midrule
A2 & 0.00031(6) & -0.0043(9) & 0.0536(11) \\
B & -0.00001(5) & -0.0004(12) & 0.0428(7) \\
\bottomrule
\end{tabular}
 \caption{Quark mass shifts to correct the mistuning of $\phi_4$ and $\phi_5$ and final lattice spacings.}
 \label{tab:shifts}
\end{table}

\section{Systematic errors}
\label{sec:sysers}

\subsection{Lattice artifacts}\label{sec:lart}
\begin{figure}[h]
   \centering
   \includegraphics[width=0.497\linewidth]{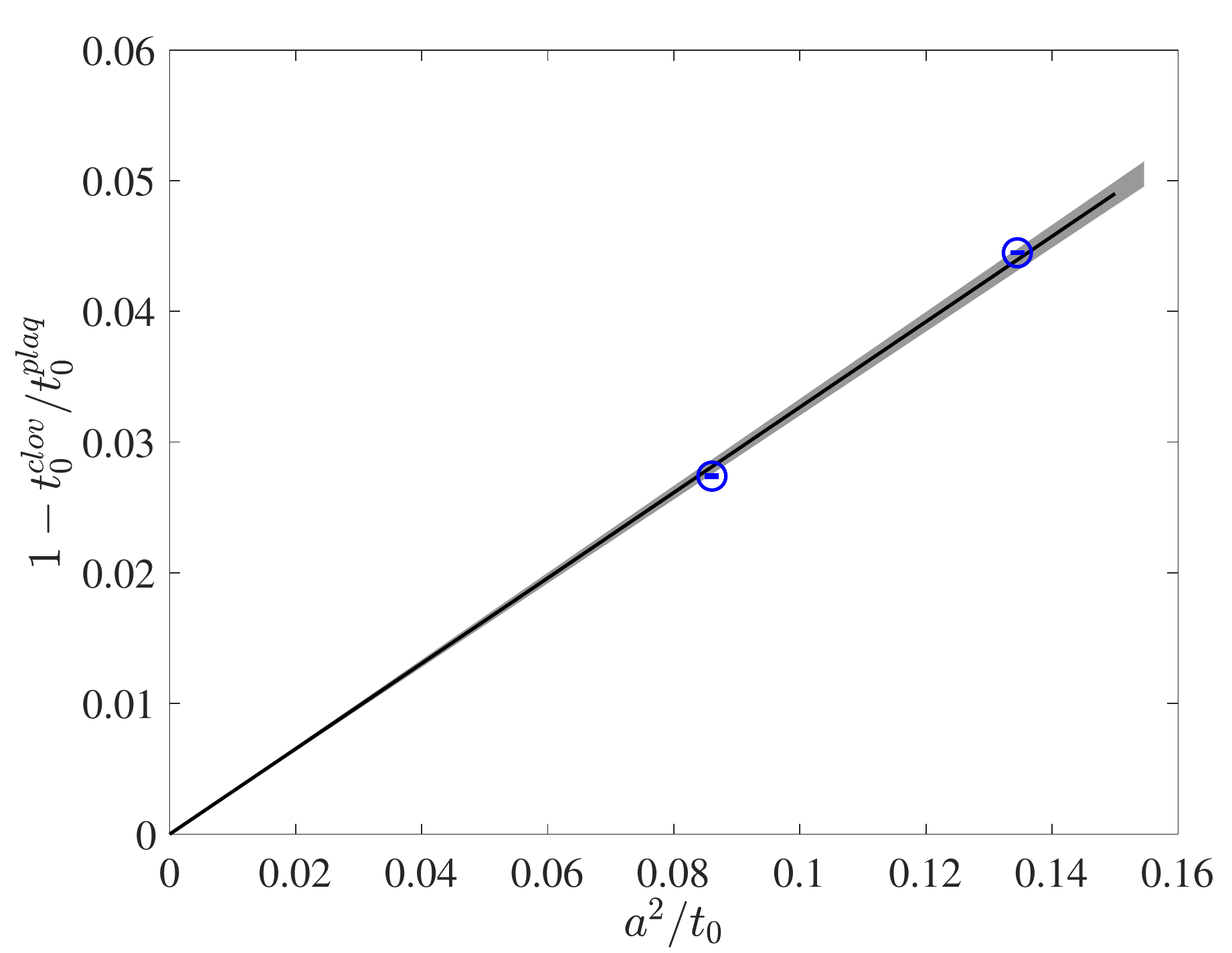}
   \includegraphics[width=0.497\linewidth]{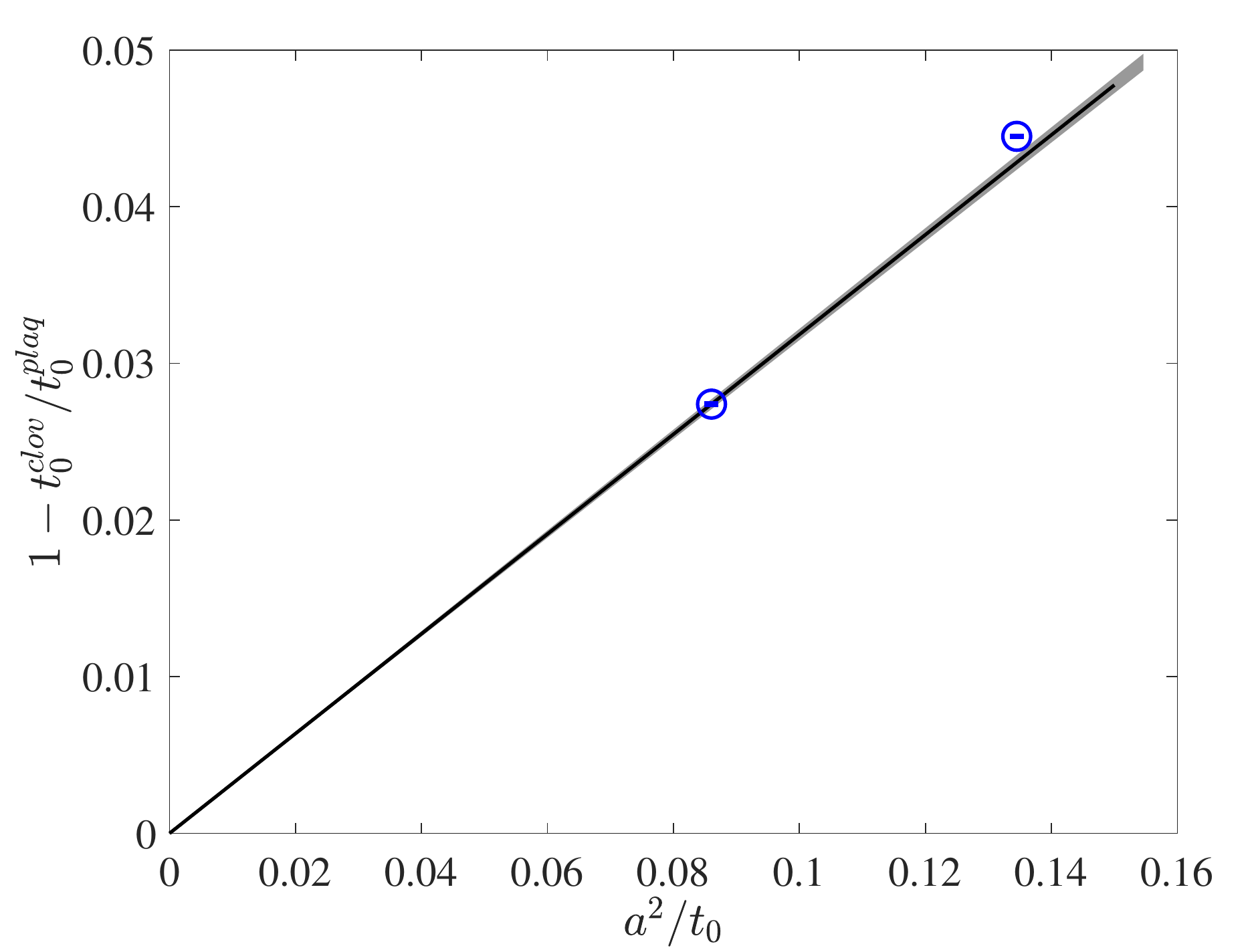}
   \caption{\label{fig:latart} The relative difference of the clover and plaquette definitions of the gradient flow scale $t_0$, which vanishes in the continuum limit. The left plot shows a linear fit in $a^2$ through our data at two lattice spacings. The right plot shows the tiny deviation from pure $a^2$ scaling through our finest lattice.
     }
\end{figure}

A major challenge in the simulation of QCD with heavy quarks, is the control of lattice artifacts. In order to 
asses their size for our ensembles, and to find out whether they behave like the expected leading scaling violations,
we look at two slightly different definitions of $t_0$ in \eq{eq:t0}, using either 
Wilson's plaquette discretization of the action density $\flowE(t)$ in \eq{eq:E}, or a more symmetric definition based on the 
symmetric ``clover'' discretization of the field strength tensor, see~\cite{Luscher:2010iy}.
Both definitions of $t_0$ have to agree in the continuum limit, but may differ at finite lattice spacing, hence, the ratio of $t_0^\mathrm{clov}$  and $t_0^\mathrm{plaq}$ has to be one up to cutoff effects. The two values of $t_0$ are highly correlated and their ratio can be evaluated very accurately. \Fig{fig:latart} shows one minus the ratios for ensembles A2 at $\beta=3.24$ and B at $\beta=3.43$ (shifted to the correct tuning point using \eq{eq:shift}), supposed to vanish in the continuum limit at a rate proportional to $a^2$.
Due to the extremely high precision of this particular observable, the data cannot be described by a pure $a^2$ function shown in the left plot of \Fig{fig:latart}, as
that would lead to an unacceptably large $\chi^2/{\rm dof}=19.2$. However, in absolute terms, the deviation from pure $a^2$ scaling
is tiny, only about 1.7 permille on the coarser lattice for this particular observable, see the right plot of \Fig{fig:latart}. 
Given that gradient flow observables in general and the plaquette definition of the action density in particular are known to have
large lattice artifacts (cf. \cite{Luscher:2010iy})
, we are convinced that for the majority of observables continuum extrapolations linear in $a^2$ will suffice.

\subsection{Finite volume effects}

\label{sec:finV}
\begin{figure}[h]
  \begin{minipage}{0.51\linewidth}
    \includegraphics[width=\linewidth]{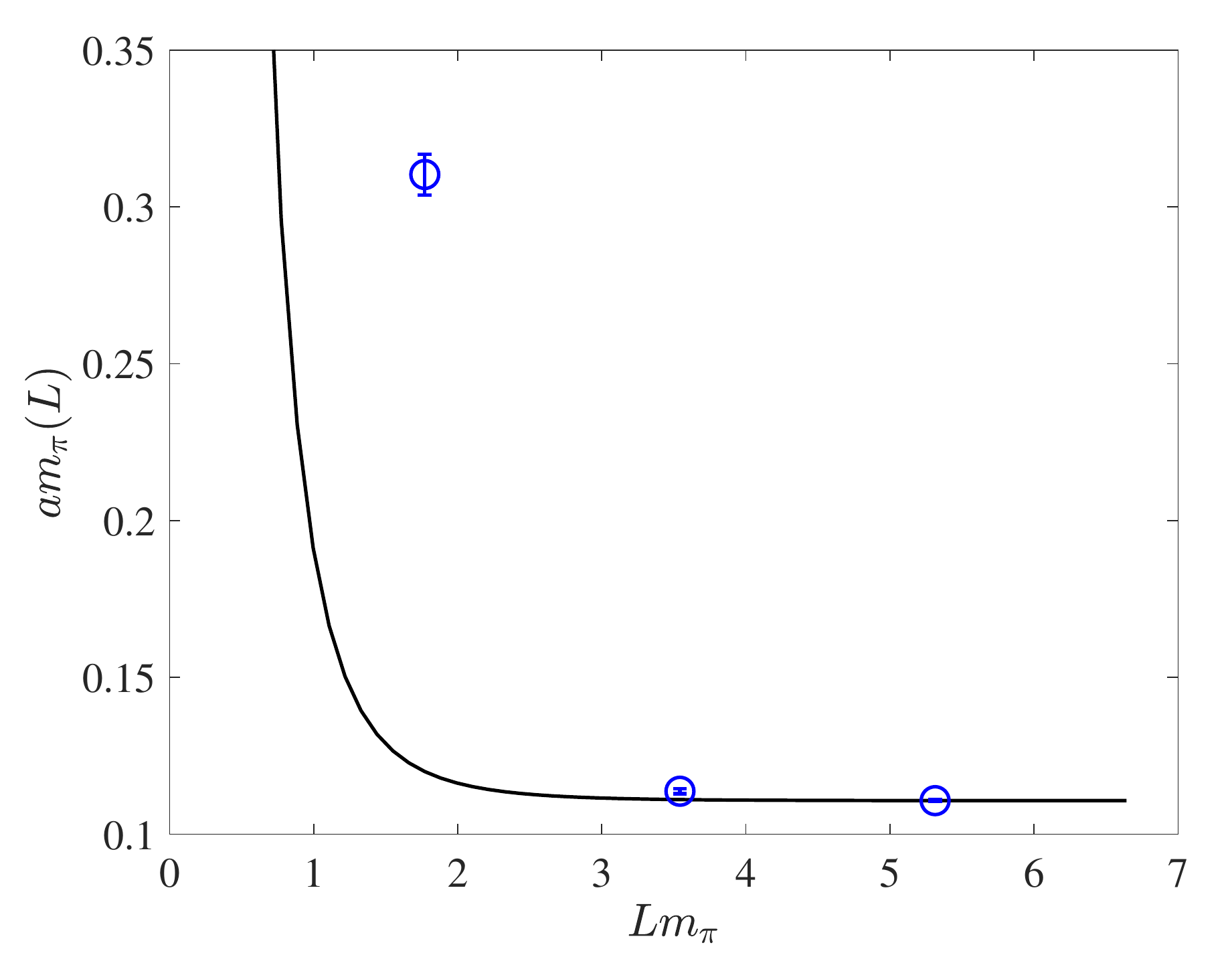}
  \end{minipage}
  \hfill
  \begin{minipage}{0.49\linewidth}
    \includegraphics[width=\linewidth]{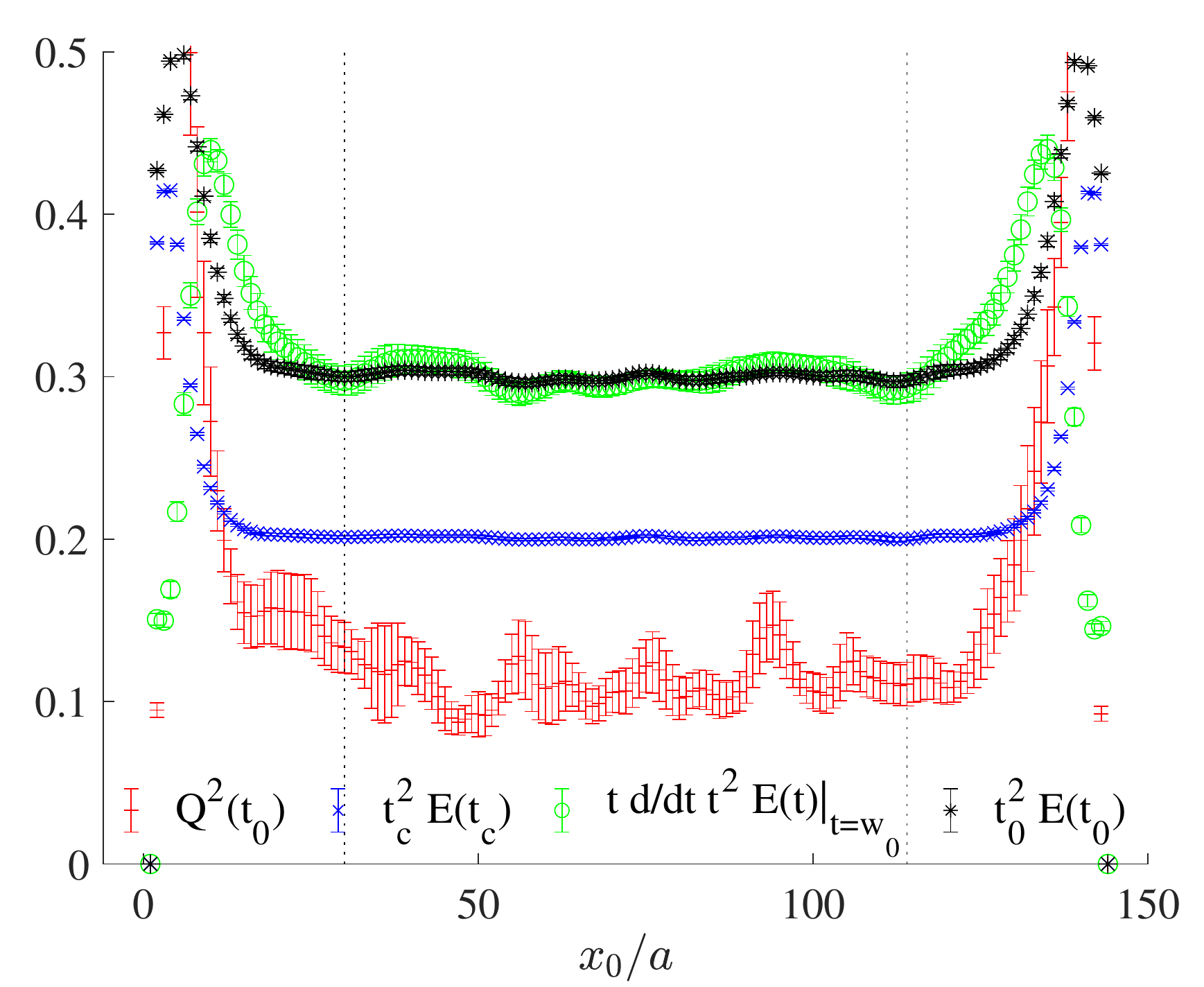}
  \end{minipage} 
  \caption{\label{fig:finvol} Left: finite volume scaling effect of $am_\pi$ represented by the analytic $\chi$PT formula in Eq.~\ref{eq:chiptfs} from \cite{Colangelo:2005gd}.
  Right: Flow measurement plateaus for $t_0^2\flowE(x_0,t_0)$ (black stars), $t_c^2\flowE(x_0,t_c)$ (blue crosses), $t\,d/dt\,t^2\flowE(x_0,t)|_{t=w_0^2}$ (green circles) and squared topological charge $Q^2(x_0,t_0)$ (scaled by $t_0^2\sum_{x_0}\flowE(x_0,t_0)/2/\sum_{x_0}Q^2(x_0,t_0)$, red plusses) of the lattice with the finest lattice spacing (ensemble B). Due to open boundary conditions we average only over time slices $x_0/a=30-114$, indicated by the vertical dotted lines for the values given in \tab{tab:meas}.}
\end{figure}

Next, we study finite volume effects of $am_\pi$ following \cite{Gasser:1986vb,Colangelo:2005gd}, who propose an analytic scaling formula from chiral perturbation theory ($\chi$PT) in the ``p-expansion"
\beq
m_\pi(L)=m_\pi(\infty)\bigg[1+\dfrac{\xi_\pi\tilde g_1(Lm_\pi(\infty))}{2\Nf}+\mathcal{O}(\xi_\pi^2)\bigg],\;\tilde g_1(x)=\sum_{n=1}^\infty\dfrac{4m(n)}{\sqrt{n}x}K_1(\sqrt{n}x),\;\xi_\pi=\dfrac{m_\pi^2}{(4\pi f_\pi)^2}.\label{eq:chiptfs}
\eeq
We take the pion mass and decay constant in $\xi_\pi$ at the SU(3) flavor symmetrical point determined on the finest lattice in~\cite{Bruno:2016plf}, $m_\pi(\infty)$ our result from ensemble A2, $\Nf=3$ and $m(n)$ are integer multiplicities listed in \cite{Colangelo:2005gd} for $n=1\ldots20$. In the left plot of Fig.~\ref{fig:finvol} we show the analytic $\chi$PT prediction together with our results on ensembles A0, A1 and A2 at the same lattice spacing. In the range of pion masses and volumes considered the agreement between the one-loop analytical prediction and our lattice data is poor, especially for $Lm_\pi<3.5$. Finite volume effects seem to be under control for $m_\pi L > 4$, as can be seen by direct comparison of ensembles A1 and A2, see e.g. \tab{tab:meas}-\ref{tab:dimless}, most values agree within errors.

Another source of finite volume effects stems from the finite temporal extent with the open boundaries.
Translational invariance is broken and only a portion of the temporal lattice sufficiently far away from
the boundaries can be used in the averages.
The right plot of Fig.~\ref{fig:finvol} shows these boundary effects for some of the flow observables
and the topological charge squared.

\subsection{Decoupling of charm}

\begin{figure}[h]
   \centering
   \includegraphics[width=\linewidth]{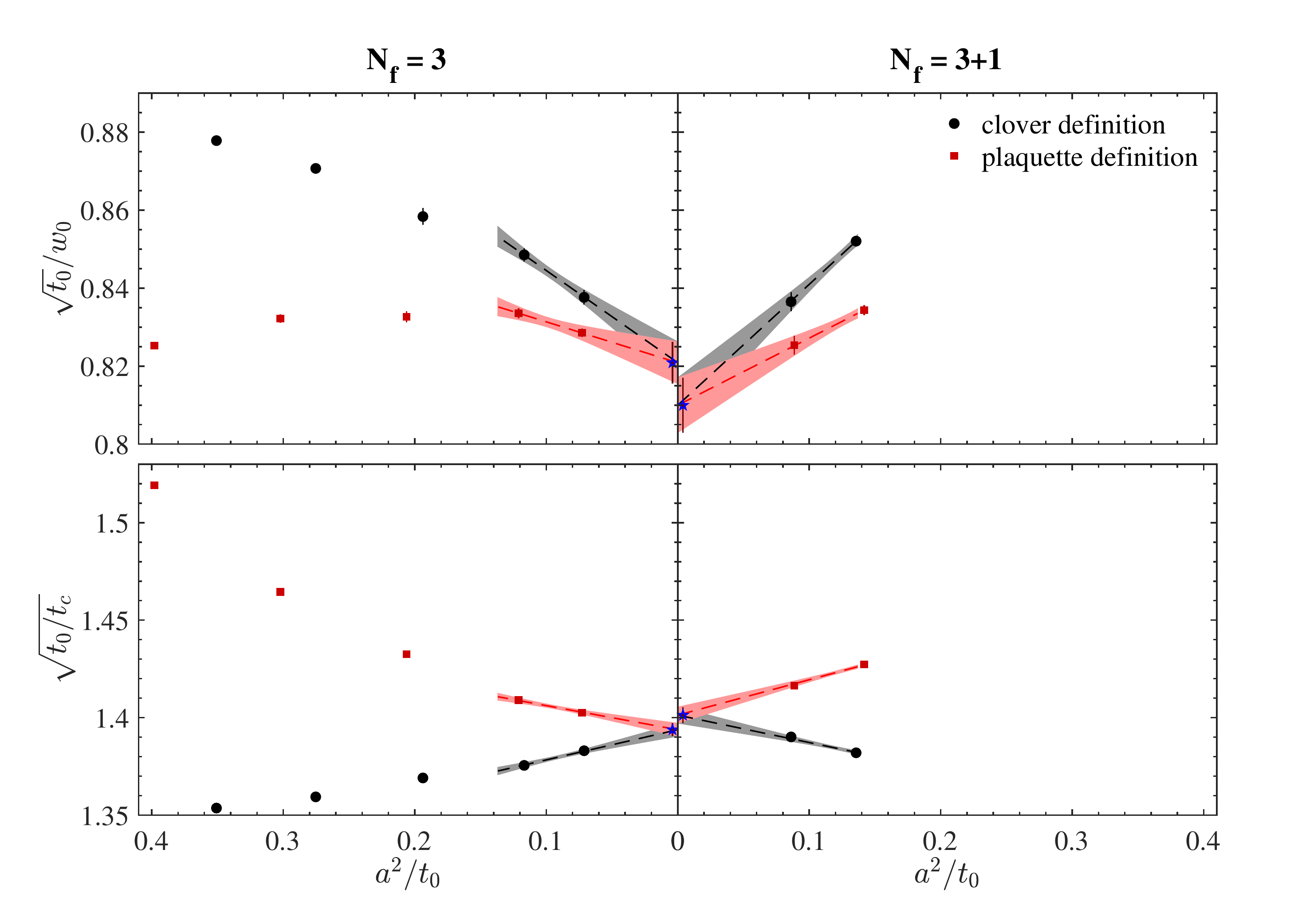}
   \caption{Comparison of continuum limit extrapolations of $\sqrt{t_0/t_c}$ (bottom) and $\sqrt{t_0/w_0^2}$ (top) of our $\Nf=3+1$ data (right) with corresponding $N_f=3$ CLS results (left) from~\cite{Bruno:2014jqa,Bruno:2016plf}
     including the finest ensemble J500, cf. \cite{Bali:2019yiy}. }
   \label{fig:cont}
\end{figure}

How well decoupling of a dynamical charm quark holds has been studied in a sequence of papers
\cite{Bruno:2014ufa,Knechtli:2017xgy,Athenodorou:2018wpk,Cali:2019enm}
Based on the outcome of these studies we are optimistic that using the $t_0^*$ value from the
$\Nf=3$ simulations introduces negligible errors for $\Nf=3+1$ QCD with a physical charm quark.
The studies were carried out in a model, namely $\Nf=2$ QCD with two degenerate charm quarks and
no light quarks. Here we are in the position to investigate decoupling in a more realistic setting with
three degenerate light quarks.
We look at dimensionless quantities from ratios of flow observables $\sqrt{t_0/t_c}$ and $\sqrt{t_0/w_0^2}$ and attempt continuum limit extrapolations from the corresponding values from ensembles A2 and B, listed in \tab{tab:dimless}. 
In~\fig{fig:cont} we plot the ratios and their extrapolations and compare with corresponding extrapolations 
on $\Nf=3$ CLS ensembles~\cite{Bruno:2014jqa}.
The ratios were formed from flow measurements with clover (black) and plaquette (red) definitions
of the action density $\flowE(t)$. The continuum extrapolations are performed by constrained
linear fits which take the correlations into account.
The deviations between $\Nf=3+1$ and $\Nf=3$ in the continuum
are far below 1\% and of the order of magnitude found in the model study \cite{Knechtli:2017xgy}.
We notice that for a reliable continuum extrapolation of the $\Nf=3$ data the combination of the
two different definitions of the action density and fine lattice spacings $\lesssim0.05$fm are
necessary.
As we already remarked in the discussion of  \fig{fig:latart} the extremely high precision
of the flow observables exposes cut-off effects which cannot be described by a pure
$a^2$ function. This is particularly visible in \fig{fig:cont} for flow observables
based on the plaquette definition of the action density.
There can be different explanations for the observed behavior of the cut-off effects.
One is large $a^4$ contributions. Another is that
while the symmetries of the clover-definition of the action density allow only for 
even powers of $a$ in the $a$-expansion of the observable, the same is not true for 
the plaquette definition with open boundary conditions~\cite{Ramos:2015baa}.
Finally
$O(a^2)$ improved variants of both the action density and the flow equations
should be employed, when possible~\cite{Ramos:2015baa} and we will investigate
these variants in the future.

\subsection{Autocorrelations}

\begin{figure}[h]
   \centering
   \includegraphics[width=0.497\linewidth]{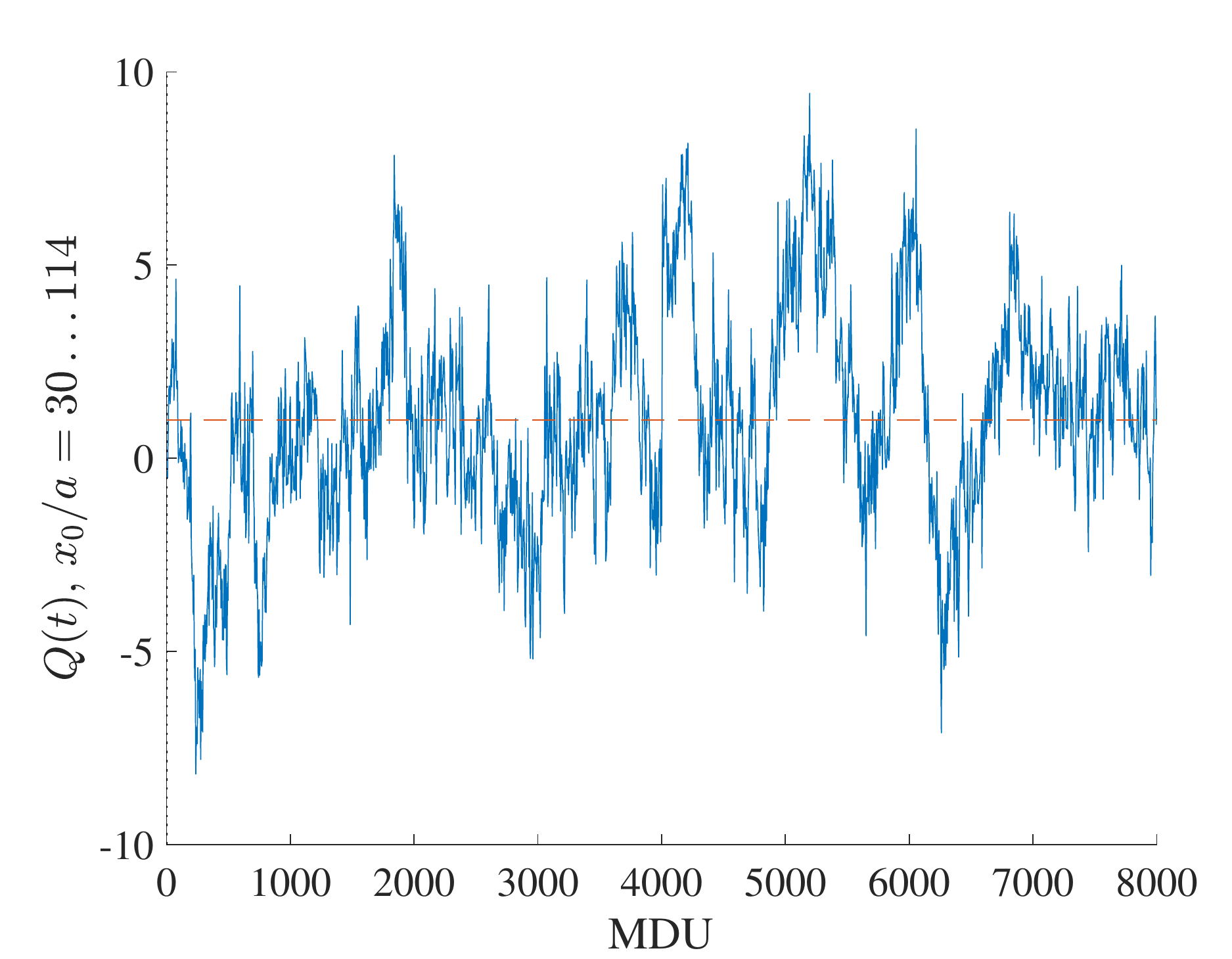} 
   \includegraphics[width=0.497\linewidth]{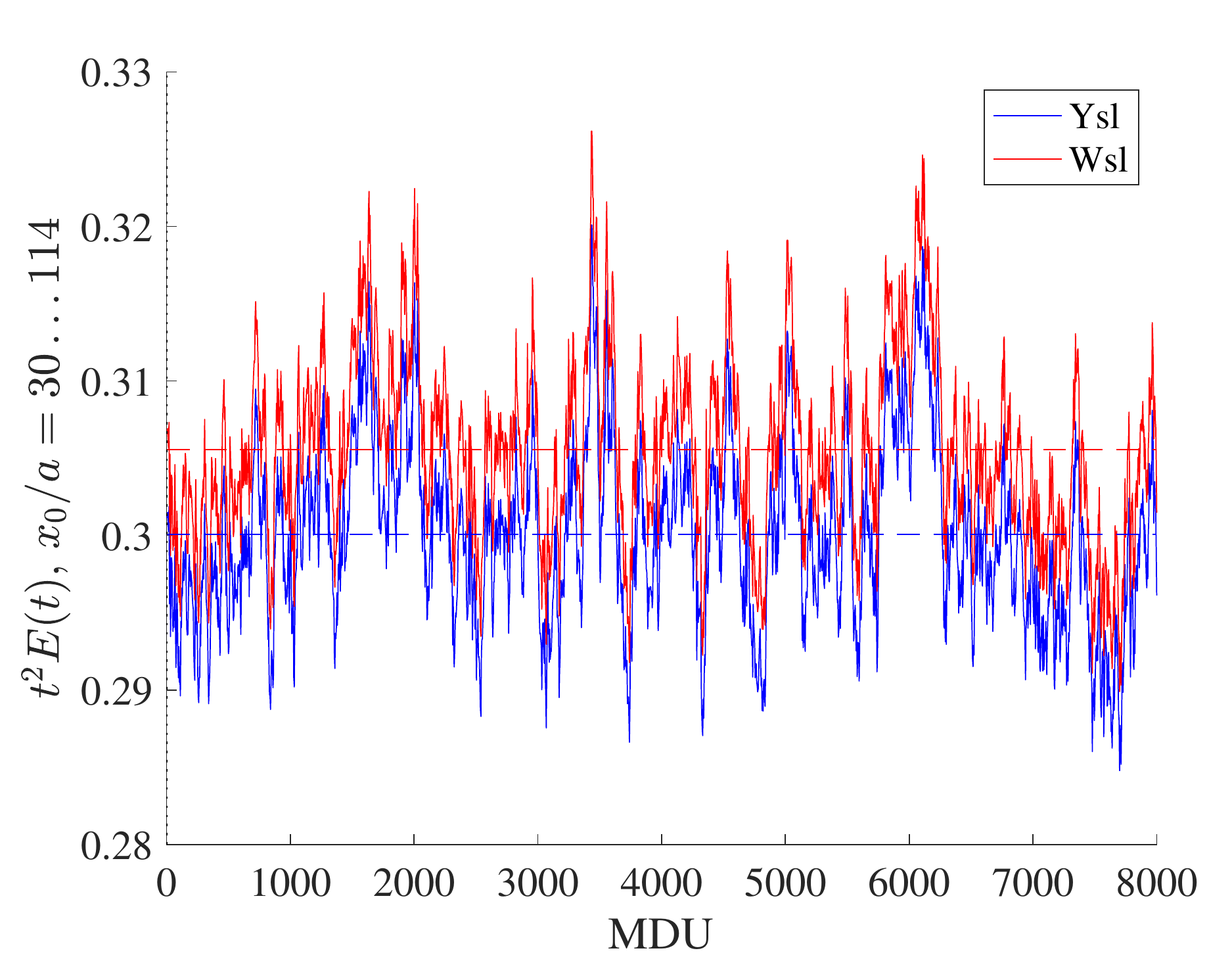}
\caption{Histories of the topological charge $Q(t)$ (left) and of $t^2\flowE(t)$ (right) of the lattice with the finest lattice spacing (both replica of ensemble B), where $t$ corresponds approximately to $t_0$. The two curves show the results for different discretizations of $\flowE(t)$, symmetric ``clover'' (blue) and plaquette (red), which differ by $O(a^2)$ effects. Due to open boundary conditions we average only over time slices $x_0/a=30\ldots114$, as indicated in the ordinate label.} 
   \label{fig:histories}
\end{figure}

When approaching small lattice spacings, the HMC algorithm is known to suffer from critical
slowing down.
Autocorrelations grow at least  $\propto a^{-2}$, but have been observed to grow much faster for
quantities like the topological charge \cite{Luscher:2011kk}.
One method to circumvent the freezing of the topological charge, is to use
open boundary conditions in one of the lattice directions \cite{Luscher:2011kk},
and that is what we are doing here.
As can be seen in Fig.~\ref{fig:histories} the topological charge moves freely, as do other ``slow'' quantities like the
flowed action density. Other quantities exhibit the expected 
slowing down $\propto a^{-2}$, which however is still manageable on our finest lattice. E.g. we find
integrated autocorrelation times $\tau_{{\rm int}, Q^2} \approx 8 \pm 3 \ \text{[4 MDU]}$ for the topological charge squared and $\tau_{{\rm int}, t_0} \approx 35 \pm 21 \ \text{[4 MDU]}$
for $t_0$.
These values are large, but small enough compared to our total statistics.
This is corroborated by fits to the tail of the autocorrelation functions which allow us to estimate the exponential autocorrelation time $\tau_{\rm exp}$ (corresponding to the slowest mode in the Markov chain) listed in Table~\ref{tab:sim}. For our error analysis we use the $\Gamma$-method \cite{Wolff:2003sm} adding a tail to the autocorrelation function to account for $\tau_{\rm exp}$ \cite{Schaefer:2010hu}. For most observables the errors are only marginally different with respect to the standard analysis
of \cite{Wolff:2003sm}.

\subsection{Chiral extrapolation effects in charmonia}\label{sec:chex}

Although our $\Nf=3+1$ ensembles are not at the physical mass point, this work also provides a formidable starting point for $N_{\rm f}=2+1+1$ simulations of QCD with improved Wilson quarks. The quantities $\phi_4$ and $\phi_5$ are chosen such, that the (renormalized) charm quark mass $\overline{m}_c$ and the sum of the degenerate (renormalized) light quark masses $\overline{m}_u+\overline{m}_d+\overline{m}_s$ are very close to their physical values. 
The physical mass point can be approached along chiral trajectories where $\overline{m}_u=\overline{m}_d$ is decreased and $\overline{m}_s$ is increased, while the trace of the quark mass matrix is kept constant, i.e. the sum of the differences to the physical masses is zero $(\Delta_{\overline{m}_u}+\Delta_{\overline{m}_d}+\Delta_{\overline{m}_s}=0)$. The chiral extrapolations are expected to be very flat for quantities that do not contain light ($u$,$d$,$s$) valence quarks. The derivatives of these quantities with respect to the light quark masses are equal to each other at the mass-symmetric point, so in the expression for the correction between symmetrical and physical mass point, the leading term vanishes, e.g. for the mass of the $\eta_c$ meson
\begin{equation}
    m^{\rm phys}_{\eta_c} = m_{\eta_c} + (\Delta_{\overline{m}_u}+\Delta_{\overline{m}_d}+\Delta_{\overline{m}_s})\frac{dm_{\eta_c}}{d\overline{m}_l} + O(\Delta^2)\, ,
\end{equation}
because $\Delta_{\overline{m}_u}=\Delta_{\overline{m}_d}=-0.5\Delta_{\overline{m}_s}$ and we only have $\mathrm{O}(\Delta^2)$ corrections.
Chiral extrapolations where the sum of the light quarks is kept constant have been successfully
employed for $\Nf=2+1$ simulations with Wilson fermions~\cite{Bietenholz:2011qq,Bruno:2014jqa}.

\section{Charmonium spectrum}
\label{sec:spectrum}

\begin{figure}
   \centering
   \includegraphics[width=1.01\linewidth]{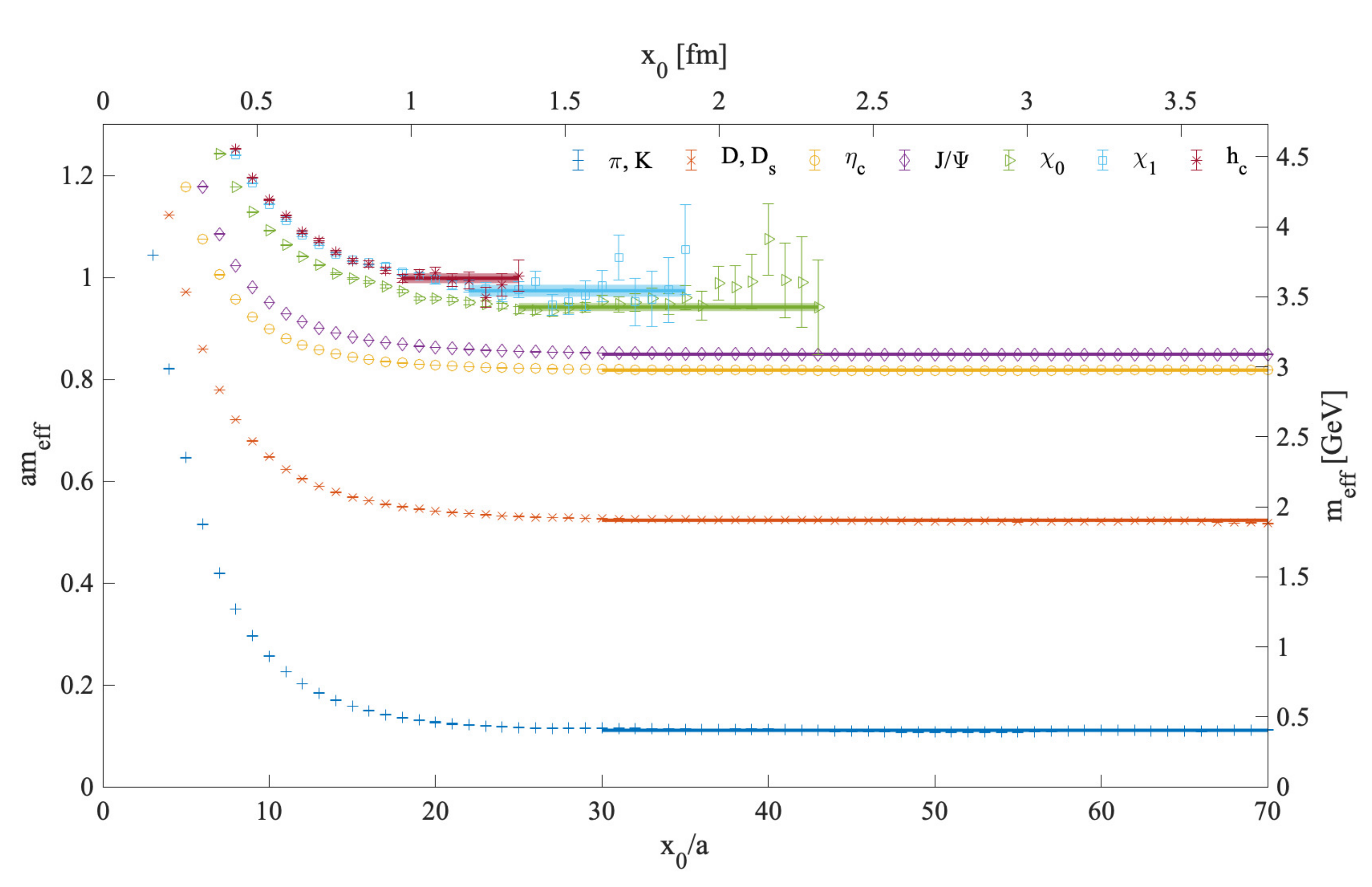}
   \includegraphics[width=1.01\linewidth]{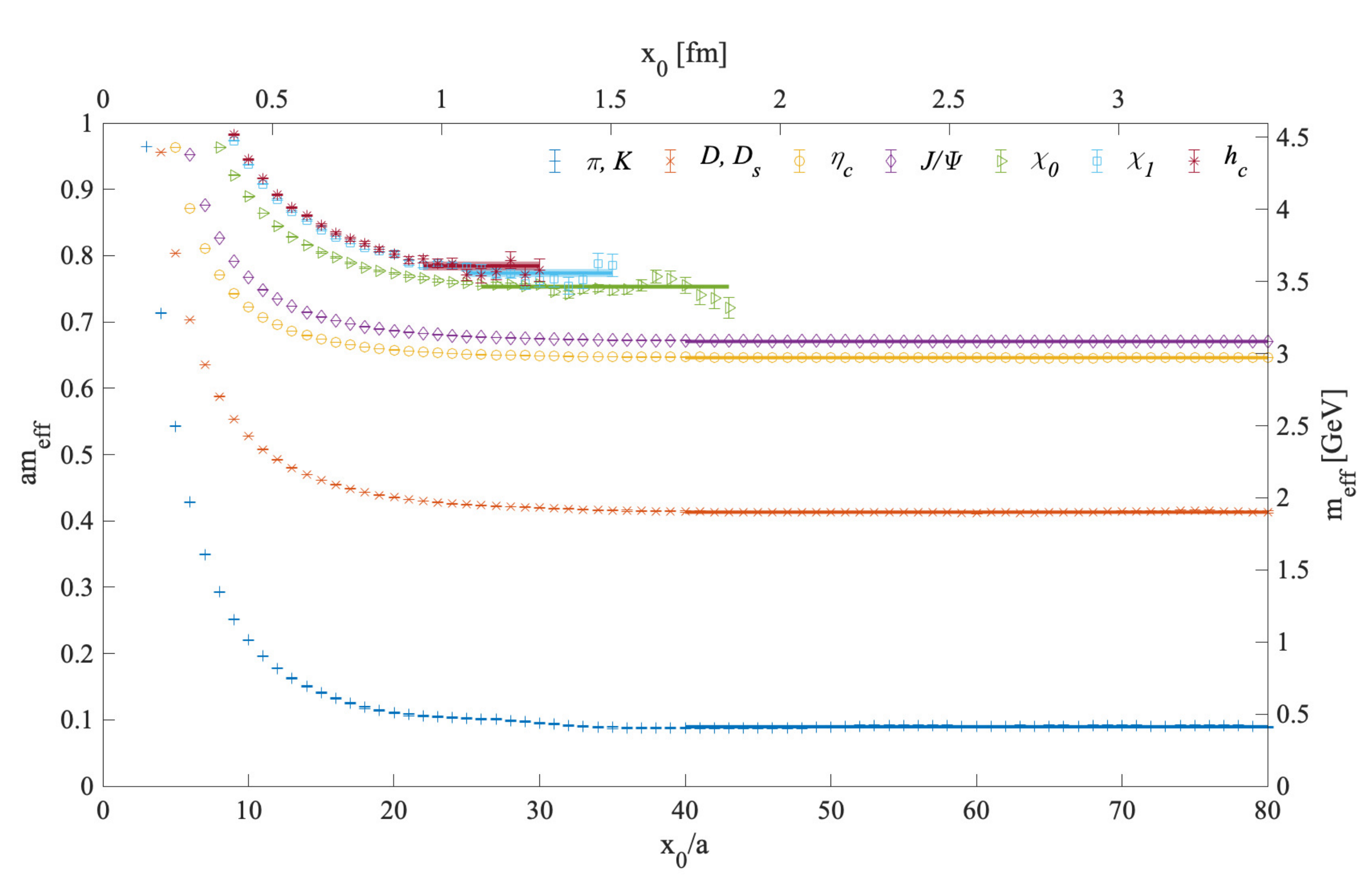}
   \caption{Effective masses of the pion/kaon, $D$- and $D_s$-meson, charmonium states $\eta_c$, $J/\Psi$, $\chi_0$, $\chi_1$ and $h_c$ (from bottom to top) on ensembles A2 (top) and B (bottom).}
   \label{fig:meff}
\end{figure}

The new ensembles with the fine lattice spacings are very well suited for a study of charmonia, already at the coarsest lattice, the charmonium masses that we measure,  neglecting disconnected contributions at the moment, are very close to their values in nature. In figure~\ref{fig:meff} we show the meson spectra of our ensembles A2 and B. We get a clear signal up to the $J/\psi$ state and can extract reasonable plateau values for higher lying states summarized in table~\ref{tab:charm}. We find good agreement with PDG~\cite{Tanabashi:2018oca} data because the states contain only charm valence quarks which in our simulations have their physical mass value. Further, we get a very precise result for the charmonium hyperfine splitting $(m_{J/\psi}-m_\eta)/m_\eta$ with perfect agreement to the experimentally known value $0.038$, cited by the Particle Data Group, see~\tab{tab:hyperfine}. 

We also evaluate a number of other dimensionless quantities from ratios of flow observables and meson masses, summarized in~\tab{tab:dimless}. For the latter and the charmonium masses in physical units we attempt continuum limit extrapolations via linear fits of the corresponding values from ensembles A2 and B. These results are also listed in tables~\ref{tab:charm}-\ref{tab:dimless}.

\begin{table}[h]
   \centering
\begin{tabular*}{\textwidth}{c @{\extracolsep{\fill}}ccccccc}
\toprule  
 & ens. & $\eta_c$ & $J/\psi$ & $\chi_{c_0}$ & $\chi_{c_1}$ & $h_c$ \\
\midrule
\multirow{5}{*}{$am_{eff}$} & A1 & 0.8175(3) & 0.8489(6) & 0.943(13) & 0.970(19) & 0.995(16) \\
\cline{2-7}
					 & A2 & 0.8180(1) & 0.8492(2) & 0.9418(94) & 0.9744(139) & 0.9983(116) \\
					 &      &  0.8114(15) & 0.8424((18) & 0.935(25) & 0.978(36) & 0.999(32) \\
\cline{2-7}
& B & 0.6461(2) & 0.6705(3) & 0.7508(50) & 0.7693(94) & 0.793(11) \\
&	& 0.6453(20) & 0.6699(19) & 0.7507(56) & 0.768(11) & 0.792(12) \\
\midrule
\multirow{7}{*}{$m_{eff}$ [GeV]} & A1 & 3.010(63) & 3.126(66) & 3.47(12) & 3.57(14) & 3.66(14) \\
\cline{2-7}
						 & A2 & 3.001(62) & 3.116(65) & 3.46(11) & 3.58(12) & 3.66(12) \\
						 &      & 2.990(67) & 3.104(70) & 3.44(16) & 3.61(21) & 3.68(19) \\
\cline{2-7}
& B & 2.973(52) & 3.086(55) & 3.456(79) & 3.54(10) & 3.65(10) \\
&	& 2.973(60) & 3.086(62) & 3.458(81)	& 3.54(100) & 3.65(11)\\
\cline{2-7}
continuum & limit & 2.93(18) & 3.03(19) & 3.45(29) & 3.47(33) & 3.63(35) \\
		  &	      & 2.94(20) & 3.05(21) & 3.49(36) & 3.41(46) & 3.60(46) \\
\midrule
PDG [GeV] & &  2.9834(5) & 3.096900(6) & 3.4148(3) & 3.51066(7) & 3.52538(11) \\
\bottomrule
\end{tabular*}
\caption{Effective lattice and physical masses of charmonium states $\eta_c$, $J/\psi$, $\chi_{c_0}$, $\chi_{c_1}$ and $h_c$ together with continuum limit extrapolations and their PDG~\cite{Tanabashi:2018oca} values, the second row values for ensembles A2 and B are shifted to the correct tuning points $\phi_4$ and $\phi_5$, for details see \sect{sec:rmd}, \eq{eq:shift}.}\label{tab:charm}
\end{table}

\begin{table}[h]
\centering
\begin{tabular}{rccc}
\toprule  
 & A2 & B & cont.\\
$(m_{J/\Psi}-m_\eta)/m_\eta$ & 0.0381(1) & 0.0378(2) & 0.0374(6)\\ 
shifted &  0.0382(3) & 0.0380(3) & 0.0376(11) \\
\bottomrule
\end{tabular}
 \caption{Charmonium hyperfine splitting $(m_{J/\Psi}-m_\eta)/m_\eta=0.038$ (PDG) on various ensembles. The last column gives the continuum limit extrapolations of original and shifted results.}
 \label{tab:hyperfine}
\end{table}

\begin{table}[h]
\centering
\begin{tabular*}{\textwidth}{l @{\extracolsep{\fill}}cccccccc}
\toprule  
ens. & $\sqrt{t_0/t_c}$ & $\sqrt{t_0/w_0^2}$ & $\dfrac{m_D}{m_\pi}$ & $\dfrac{m_\eta}{m_\pi}$ & $\dfrac{m_{J/\Psi}}{m_\pi}$ & $\dfrac{m_{\chi_{c_0}}}{m_\pi}$ & $\dfrac{m_{\chi_{c_1}}}{m_\pi}$ & $\dfrac{m_{h_c}}{m_\pi}$ \\
\midrule
A1 & 1.3836(16) & 0.851(3) & 4.58(4) & 7.16(7) & 7.44(7) & 8.26(13) & 8.49(18) & 8.71(16) \\ 
\midrule
A2  & 1.3820(8) & 0.852(1) & 4.71(1) & 7.36(2) & 7.64(2) & 8.48(9) & 8.77(13) & 8.99(11) \\ 
    & 1.3827(9) & 0.851(2) & 4.6267 & 7.27(1) & 7.55(1) & 8.38(23) & 8.77(33) & 8.96(29) \\
\midrule
B & 1.3898(15) & 0.837(2) & 4.62(2) & 7.21(4) & 7.49(4) &  8.38(6) & 8.59(9) & 8.86(11) \\
   & 1.3900(21) & 0.837(4) & 4.6267      & 7.23(1) & 7.51(1) & 8.41(6) & 8.61(10) & 8.88(11) \\
\midrule
cont. & 1.4043(37) & 0.809(7) & 4.45(6) & 6.95(10) & 7.21(11) & 8.22(23) & 8.27(33) & 8.63(35) \\
	& 1.4043(48) & 0.809(9) & 4.6267    & 7.16(3) & 7.43(4) & 8.47(43) & 8.32(64) & 8.74(60) \\
\bottomrule
\end{tabular*}
 \caption{Dimensionless quantities on various ensembles, the second row values for ensembles A2 and B are shifted to the correct tuning points $\phi_4$ and $\phi_5$, using the strategy described in \sect{sec:rmd}, \eq{eq:shift}. Note, there is no error for the shifted $m_D/m_\pi$ ratio, since it is given by the exact ratio of $\phi_5/\sqrt{6\phi_4}$. The last two lines correspond to continuum limit extrapolations of original and shifted values via linear fits of corresponding values from ensembles A2 and B.}
 \label{tab:dimless}
\end{table}

\section{Conclusions \& outlook}
\label{sec:concl}

We presented the scale setting and tuning of $\Nf=3+1$ QCD using a massive renormalization scheme with a non-perturbatively determined clover coefficient from~\cite{Fritzsch:2018kjg}. We produced two ensembles A1 and A2 with lattice sizes $96\times32^3$ and $128\times48^3$ at a lattice spacing $a=0.054$ fm, see also~\cite{Hollwieser:2019kuc}, and an ensemble B on a $144\times48^3$ lattice at a finer lattice spacing $a=0.043$fm.

As a first physics result, we measure the masses of the charmonium states $\eta_c$, $J/\psi$, $\chi_{c_0}$, $\chi_{c_1}$ and $h_c$, which agree with their PDG values. In particular, we can reproduce the charmonium hyperfine splitting $(m_{J/\Psi}-m_\eta)/m_\eta$ within permille level precision of the experimentally known value $0.038$. First continuum limit extrapolations of the measured quantities are attempted.

The next steps are to double the statistics of ensemble B and produce a third ensemble C on a $192\times64^3$ lattice at an even finer lattice spacing $a=0.032$fm, in order to make more reliable continuum limit extrapolations.
For the future, we plan to measure disconnected contributions to the charmonium masses and extract excited states.
A further development of this project is the simulation of $\Nf=2+1+1$ QCD close to the physical light quark masses. Another longterm goal is the determination of the strong coupling $\alpha_S$ or equivalently the $\Lambda$-parameter in the case of four flavors. 
The currently most precise result comes from lattice simulations~\cite{Bruno:2017gxd},
see also~\cite{Aoki:2019cca}.
One of the remaining uncertainty comes from the use of perturbative decoupling for the
matching of the gauge couplings of QCD with four and three flavors.
In \cite{Athenodorou:2018wpk} this uncertainty was estimated to be below 1.5\% for the
ratios of the $\Lambda$ parameters which is still below the 3.5\% precision of the $\Lambda$-parameter of~\cite{Bruno:2017gxd}.
But new techniques like~\cite{DallaBrida:2019mqg} promise a larger precision for $\alpha_S$. The running of $\alpha_S$ in four flavor QCD has been computed in~\cite{Tekin:2010mm}. 
Such a result, can be combined with the scale setting for $\Nf=3+1$ QCD presented in this work to obtain $\alpha_S$ fully non-perturbatively in four flavor QCD.

\section*{Acknowledgements}

We thank Rainer Sommer and Ulli Wolff for precious advice on the manuscript and
Andrei Alexandru, Stefan D{\"u}rr and Stefan Schaefer for valuable discussions.
We are grateful to our CLS colleagues for sharing data used in Fig.~\ref{fig:cont}.
We gratefully acknowledge the Gauss Center for Supercomputing e.V. (\url{www.gauss-centre.eu}) for funding this project by providing computing time on the GCS Supercomputer JUWELS at J{\"u}lich Supercomputing Centre (JSC) under GCS/NIC project ID HWU35. R.H. was supported by the Deutsche Forschungsgemeinschaft in the SFB/TRR55.

\begin{appendices}
\numberwithin{equation}{subsection}

\clearpage

\section{Simulation parameters}

\begin{table}[h!]
\begin{tabular}{l l l l}
\toprule
ensemble      & A1           & A2           & B          \\
ID		& nf31H100 & nf31N200 & nf31I300 \\
\midrule
Force 0, lvl  & gauge, 0       & gauge, 0       & gauge, 0    \\
Force 1, lvl  & TM1-EO-SDET, 1$\qquad$ & TM1-EO-SDET, 1$\qquad$ & TM1-EO-SDET, 1 \\
Force 2, lvl  & TM2-EO, 1      & TM2-EO, 1      & TM2-EO, 1    \\
Force 3, lvl  & TM2-EO, 1      & TM2-EO, 1      & TM2-EO, 1      \\
Force 4, lvl  & TM2-EO, 1      & TM2-EO, 1      & TM2-EO, 1      \\
Force 5, lvl  & TM2-EO, 2      & TM2-EO, 2      & TM2-EO, 2      \\
Force 6, lvl  & s-RAT-SDET, 1  & s-RAT-SDET, 1  & s-RAT-SDET, 1\\
Force 7, lvl  & s-RAT, 1       & s-RAT, 1       & s-RAT, 1\\
Force 8, lvl  & s-RAT, 1       & s-RAT, 1       & s-RAT, 1    \\
Force 9, lvl  & s-RAT, 2       & s-RAT, 2       & s-RAT, 2      \\
Force 10, lvl & s-RAT, 2       & s-RAT, 2       & s-RAT, 2    \\
Force 11, lvl & s-RAT, 2       & s-RAT, 2       & s-RAT, 2      \\
Force 12, lvl & s-RAT, 2       & s-RAT, 2       & s-RAT, 2      \\
Force 13, lvl & c-RAT-SDET, 1  & c-RAT-SDET, 1  & c-RAT-SDET, 1 \\
\midrule
Level 0,nstep & OMF4, 2        & OMF4, 2        & OMF4, 2    \\
Level 1,nstep & OMF4, 1        & OMF4, 1        & OMF4, 1       \\
Level 2,nstep & OMF2, 8        & OMF2, 8        & OMF2, 8       \\
\midrule 
$\kappa_{uds}$& 0.13440733       & 0.13440733       & 0.135990    \\
$\kappa_c$    & 0.127840       & 0.127840       & 0.130880    \\
$c_{sw}$ & 2.18859 & 2.18859 & 1.914633         \\
$a\mu_0$    & 0.0005 & 0.0005 & 0.0005 \\
$a\mu_1$    & 0.005 & 0.005 & 0.005 \\
$a\mu_2$    & 0.05 & 0.05 & 0.05 \\
$a\mu_3$    & 0.5 & 0.5 & 0.5 \\
$N_p^s, [r_a,r_b]^s$ & 12, [0.001,9.0] & 12, [0.001,9.0] & 12, [0.002,8.0]     \\
$N_p^c, [r_a,r_b]^c\qquad\qquad$ & 10, [0.2,8.0] & 10, [0.2,8.0] & 8, [0.2,8.0] \\
\midrule
$N_{\rm traj}$& 3908             & 3868             & 4000        \\
$\langle P_{acc}\rangle$     & 97.7\%           & 93.5\%           & 97.7\% \\
\bottomrule
\end{tabular}
\caption{Parameters of the algorithm: We give the forces used for the gauge and fermion fields with their integration levels, the integrators for the different levels and number of steps per trajectory resp. outer level, the hopping, $c_{sw}$ and twisted-mass parameters, the number of poles $N_\mathrm{p}$ and the ranges used in the RHMC for strange and charm quarks, as well as the total length of the Markov chain and the acceptance rates.}\label{tab:param}
\end{table}

\clearpage

\section{Mass derivatives}
\begin{table}[h!]
   \centering
\begin{tabular*}{\textwidth}{c @{\extracolsep{\fill}}ccccc}
\toprule  
	& $\Ob$ & $d\Ob/dm_u$ & $d\Ob/dm_d$ & $d\Ob/dm_s$ & $d\Ob/dm_c$\\
\midrule
$am_{\pi}$ & 0.1111(3) & 7.88(35) & 7.88(35) & 2.77(36) & 1.25(18)\\
$am_{K}$ & 0.1111(3) & 7.88(35) & 2.77(36) & 7.88(35) & 1.25(18)\\
$am_{D}$ & 0.5234(3) & 3.71(38) & 1.95(38) & 1.95(38) & 2.29(18)\\
$am_{D_s}$ & 0.5234(3) & 1.95(38) & 1.95(38) & 3.71(38) & 2.29(18)\\
$am_{\eta_c}$ & 0.8180(1) & 1.13(13) & 1.13(13) & 1.13(13) & 1.80(5)\\
$am_{J/\Psi}$ & 0.8492(2) & 1.51(21) & 1.51(21) & 1.51(21) & 1.92(9)\\
$am_{\chi_{c_0}}$ & 0.9418(51) & 0.9(8.3) & 0.9(8.3) & 0.9(8.3) & 1.8(4.7)\\
$am_{\chi_{c_1}}$ & 0.9744(76) & 1.2(12.3) & 1.2(12.3) & 1.2(12.3) & -0.7(7.0)\\
$am_{h_c}$ & 0.9983(62) & -23.6(10.4) & -23.6(10.4) & -23.6(10.4) & -5.5(5.6)\\
$t_0/a^2$ & 7.371(24) & -104.5(16.5) & -104.5(16.5) &-104.5(16.5) & -37.8(7.2) \\
$t_c/a^2$ & 3.860(8) & -38.0(5.8) & -38.0(5.8) & -38.0(5.8) & -15.2(2.6) \\
$w_0/a^2$ & 10.154(61) & -250.5(41.7) & -250.5(41.7) & -250.5(41.7) & -81.9(17.1) \\
$Q$ & 6.55(16) & 80.2(153.6) & 80.2(153.6) & 80.2(153.6) & -34.7(67.9) \\
$\phi_4$ & 1.092(5) & 139.4(5.8) & 72.4(5.8) & 105.9(5.8) & 18.9(3.0) \\
$\phi_5$ & 12.058(17) & -13.5(12.3) & -40.4(12.2) & -27.0(12.3) & 21.8(5.8) \\
\bottomrule
\end{tabular*}
\caption{Derivatives with respect to quark masses of pion/kaon, $D$- and $D_s$-mesons and charmonium states $\eta_c$, $J/\psi$, $\chi_{c_0}$, $\chi_{c_1}$ and $h_c$ effective masses and flow observables on ensemble A2.}\label{tab:sigmA2}
\end{table}

\begin{table}[h!]
   \centering
\begin{tabular*}{\textwidth}{c @{\extracolsep{\fill}}ccccc}
\toprule  
	& $\Ob$ & $d\Ob/dm_u$ & $d\Ob/dm_d$ & $d\Ob/dm_s$ & $d\Ob/dm_c$\\
\midrule
$am_\pi$ & 0.0896(4) & 6.72(45) & 6.72(45) & 1.46(46) & 0.82(29) \\
$am_K$ & 0.0896(4) & 6.72(45) & 1.46(46) & 6.72(45) & 0.82(29) \\
$am_D$ & 0.4135(5) & 2.42(61) & 0.57(59) & 0.57(59) & 1.72(40) \\
$am_{D_s}$ & 0.4135(5) & 0.57(59) & 0.57(59) & 2.42(61) & 1.72(40) \\
$am_{\eta_c}$ & 0.6461(1) & 0.52(12) & 0.52(12) & 0.52(12) & 1.65(8) \\
$am_{J/\Psi}$ & 0.6705(2) & 0.47(18) & 0.47(18) & 0.47(18) & 1.52(11) \\
$am_{\chi_{c_0}}$ & 0.7508(21) & -1.9(3.2) & -1.9(3.2) & -1.9(3.2) & 0.3(2.2) \\
$am_{\chi_{c_1}}$ & 0.7693(39) & -2.9(5.6) & -2.9(5.6) & -2.9(5.6) & 2.9(3.7) \\
$am_{h_c}$ & 0.7930(45) & 1.5(5.8) & 1.5(5.8) & 1.5(5.8) & 1.5(3.8) \\
$t_0/a^2$ & 11.598(55) & -140.5(40.0) & -140.5(40.0) & -140.5(40.0) & -54.8(26.2) \\
$t_c/a^2$ & 6.002(18) & -45.4(10.1) & -45.4(10.1) & -45.4(10.1) & -21.2(7.4) \\
$w_0/a^2$ & 16.574(164) & -415.7(127.2) & -415.7(127.2) & -415.7(127.2) & -137.9(79.7) \\
$Q$ & 1.63(6) & 112.4(44.0) & 112.4(44.0)  & 112.4(44.0) & 48.4(28.2) \\
$\phi_4$ & 1.116(10) & 154.0(10.4) & 154.0(10.4) & 154.0(10.4) & 15.1(6.2) \\
$\phi_5$ & 11.950(24) & -38.1(23.5) & -38.1(23.5) & -38.1(23.5) & 21.6(15.5) \\
\bottomrule
\end{tabular*}
\caption{Derivatives with respect to quark masses of pion/kaon, $D$- and $D_s$-mesons and charmonium states $\eta_c$, $J/\psi$, $\chi_{c_0}$, $\chi_{c_1}$ and $h_c$ effective masses and flow observables on ensemble B.}\label{tab:sigmaB}
\end{table}

\end{appendices}

\clearpage

\addcontentsline{toc}{section}{References}
\bibliographystyle{utphys}
\bibliography{nf3p1}

\end{document}